\documentclass[11pt,a4paper,reqno]{amsart} 
\usepackage{tikz}
\usetikzlibrary{positioning,decorations.pathmorphing}
\usepackage{wasysym}

\usepackage{amssymb,graphicx}

\usepackage{amssymb}

\def\et{\mbox{\dh}}
\def\th{\mbox{\thorn}}

\newcommand{\koniec}{\begin{flushright}  $\Box $ \end{flushright}}
\newtheorem{theo}{Theorem}[section] 
\newtheorem{prop}[theo]{Proposition}  

\newtheorem{defi}[theo]{Definition}
\newtheorem{col}[theo]{Corollary}

\def\theequation{\thesection.\arabic{equation}}

\newcounter{mnotecount}[section]

\renewcommand{\themnotecount}{\thesection.\arabic{mnotecount}}

\newcommand{\mnote}[1]
{\protect{\stepcounter{mnotecount}}$^{\mbox{\footnotesize
$
\bullet$\themnotecount}}$ \marginpar{
\raggedright\tiny\em
$\!\!\!\!\!\!\,\bullet$\themnotecount: #1} }

\newcommand{\CP}{\mathbb{CP}}

\newcommand{\HHH}{\mathbb{H}}

\newcommand{\II}{{\mathcal I}}
\newcommand{\R}{\mathbb{R}}

\def\p{\partial}

\def\be{\begin{equation}}

\def\ee{\end{equation}}

\def\bea{\begin{eqnarray}}
\def\eea{\end{eqnarray}}

\def\T{{\bf T}}

\begin{document}\date{1 March 2019}
\vspace*{-1.0cm}
\title{Conformally  isometric embeddings and Hawking temperature}
\author{Maciej Dunajski}
\address{Department of Applied Mathematics and Theoretical Physics\\ 
University of Cambridge\\ Wilberforce Road, Cambridge CB3 0WA, UK.}
\email{m.dunajski@damtp.cam.ac.uk}

\author{Paul Tod}
\address{The Mathematical Institute\\
Oxford University\\
Woodstock Road, Oxford OX2 6GG\\ UK.
}
\email{tod@maths.ox.ac.uk}
\maketitle

\begin{abstract} 
We find necessary and sufficient conditions for existence of a
locally isometric embedding of a vacuum space-time into a conformally-flat 5-space. 
We explicitly construct  such embeddings for any spherically symmetric Lorentzian metric in $3+1$ dimensions as a hypersurface in 
$\R^{4, 1}$. For the Schwarzschild metric the embedding is global, and extends through the horizon all the way to the $r=0$ singularity. We discuss the asymptotic properties of the embedding in the context of Penrose's theorem
on Schwarzschild  causality. We finally show that the Hawking temperature of the Schwarzschild metric agrees with the Unruh
temperature measured by an observer moving along hyperbolae in $\R^{4, 1}$.

\end{abstract}   
\maketitle
\section{Introduction}
The modern point of view on space--times in General Relativity is intrinsic:
a space--time is an abstract manifold with a Lorentzian metric, and neither
the topological structure nor the curvature properties invoke a notion of an 
ambient space. On the other hand an intuitive, visual representation
of  curvature is that of a surface inside a flat $\R^N$. This extrinsic approach can be put on firm mathematical footings. The Whitney embedding theorem
\cite{whitney}
 states that any $n$--dimensional smooth manifold can be embedded 
in $\R^{N}$ as a surface, where $N$ is at most $2n+1$. If $\R^N$ is equipped
with a flat pseudo--Riemannian metric, and the embedding is isometric and
global, then the upper bound on dimension is much higher. The Clarke
embedding theorem \cite{clarke} states that a smooth $n$--manifold with 
a Lorentzian metric can  be embedded isometrically in $\R^{q, p}$ where $p$ is at most 
$2$ and $q$ is at most $n(2n^2+37)/6+5n^2/2+1$.
If the isometric embedding is only required to be local then the upper bound, in the real analytic category, $N$ is at most $n(n+1)/2$ (see \cite{cartan}).

Thus, for  $3+1$ dimensional Lorentzian space--times the upper 
bound of a global isometric embedding is $89$, which puts the whole programme
outside the scope of practical considerations. Fortunately many known
exact solutions to the Einstein equations can be embedded in lower dimensions.
While it is impossible to embed the Schwarzschild solution in five (flat) dimensions, there exist several local isometric embeddings in $4+2$
dimensions, as well as a global embedding in $5+1$ dimensions due to 
Fronsdal \cite{fronsdal}. One says that the Schwarzschild metric has \emph{embedding class 2} because it locally isometrically embeds with codimension 2 in flat space.

In this paper we study  conformally isometric embeddings,
i.e. immersions of space times $(M, g)$ in a flat $\R^{q, p}$, such that
the pull back of the flat Lorentzian metric from the ambient space is in the conformal class of $g$. For local conformal embeddings
the upper bound on the dimension of the ambient
space is one less than for isometric embeddings \cite{jacobowitz}, 
and we shall show (Proposition \ref{prop1} in \S\ref{section1}) 
that the Schwarzschild metric
can be conformally embedded in $\R^{4, 1}$ and the embedding goes through the 
black-hole
horizon all the way to the $r=0$ singularity. We may say that the Schwarzschild metric has \emph{conformal embedding class 1}. The question of conformal embeddings of the Schwarzschild metric was discussed in Penrose's research group in Oxford in the late 1970's, and examples were found. However there does not seem to be any published literature from the discussions at that time (though there is the general result in \cite{jacobowitz}) but there are suggestions of it in \cite{p5}.

We begin in \S \ref{ss2} by discussing the theory of conformal embedding class 1. We find necessary and sufficient conditions (Proposition \ref{prop0}) for a vacuum metric to have conformal embedding class 1. In \S\ref{subalg}
we find some algebraic obstructions on the Weyl tensor for the existence of conformal embedding class 1. In 
particular we can rule out the existence of class 1 conformal embeddings for the Kerr metric.
Then applying Proposition \ref{prop0} to the Schwarzschild metric we show that any conformal embedding which is globally defined on a sphere of symmetry must actually be spherically-symmetric
(Theorem \ref{theop}). This clears the way for restricting consideration to spherically-symmetric 
conformal embeddings and these are discussed in \S\ref{section1} and  \S\ref{section4}.

In applications to physics conformal embeddings may be useful if the causal structure of
space--time needs to be examined. This is the case for a lot of classical, and some quantum physics.  In \S\ref{seccs} we shall show that the extension of the
Schwarzschild conformal embedding to the compactified space--time
maps past and future null infinities in $3+1$ dimensions to single points
on past and future null infinities in flat $4+1$ dimensions, and 
in \S\ref{seccal}
we discuss
this in the context of Penrose's theorem \cite{roger_paper} on Schwarzschild causality.
In \S\ref{htou} we shall argue  that the Hawking temperature is a conformally invariant
notion, and agrees with the Unruh temperature measured by an observer moving 
along  hyperbolae in $\R^{4, 1}$. An analogous observation for the isometric embeddings in $\R^{5, 1}$ was made in \cite{deser} and further developed in \cite{pestun}.
In \S\ref{time_emb} we shall construct some time--dependent conformal embeddings of the extreme 
Reissner--Nordstr\"om
black hole (they fail to be global). In Appendix A we shall prove Proposition
\ref{prop_real}, and show that the reality of the scalar invariants
$I, J$ for the Weyl spinor are the necessary and sufficient
conditions for the existence of class 1 conformal embedding 
resulting from the Gauss equation (there are other differential constraints
resulting from the Codazzi equation).
In Appendix B we review the GHP formalism used in the  proof of Theorem \ref{theop}.
In 
Appendix C we shall show that the parabolic isometric embedding of
Fujitani, Ikeda and Matsumoto \cite{japanese} can be obtained as an infinite boost
of the Kasner embedding. In Appendix D we shall relate
our conformal embedding in $\R^{4, 1}$ to the Fronsdal embedding in $\R^{5, 1}$.

Throughout the paper we shall follow the curvature conventions of \cite{pr}. As we make use of the two--component spinor
calculus, the signature of metrics in four--dimensions will be $(1, 3)$. The indices $a, b, \dots $ run from $0$ to $3$, and indices $\alpha, \beta, \dots $ on $\R^5$ run from $0$ to $4$.
\subsection*{Acknowledgements.} We are grateful 
to Gary Gibbons and Paul Townsend  for discussion about 
the Hawking temperature. The work of M.D. has been partially supported by STFC consolidated grant no. ST/P000681/1. 
\section{The theory of conformal embedding class 1}\label{ss2}
Let $\R^{r, s}$ be an $(r+s)$--dimensional vector space equipped with a metric
\[
\eta=\eta_{\alpha\beta} dX^{\alpha} dX^{\beta}, \quad \alpha, \beta =1, \dots, N=r+s
\]
of signature $(r, s)$. 
\begin{defi}
A conformally isometric embedding of a pseudo--Riemannian $n$--dimensional manifold
$(M, g)$ as a surface in $\R^{r, s}$ is a map $\iota:M\rightarrow\R^{r, s}$ such that
$\iota^*(\eta)=\Omega^2 g$ for some  $\Omega: M\rightarrow \R^+$ and
$\iota(M)\subset \R^{r, s}$ is diffeomorphic to $M$. 
\end{defi}

A theorem of Jacobowitz and Moore \cite{jacobowitz} implies that real analytic (pseudo) Riemannian
manifold of dimension $n$ can be locally conformally embedded in $\R^{r, s}$, where
$r+s=n(n+1)/2-1$. To understand this number, consider the image of $M$ in $\R^{r, s}$ in terms of embedding functions 
$X^{\alpha}=X^{\alpha}(x^a)$, where $x^a, a=1, \dots, n$ are local coordinates on $M$ such that 
$g=g_{ab}dx^adx^b$. The conformal embedding condition
becomes a system of $n(n+1)/2$ PDEs 
\be
\label{system_conf1}
\eta_{\alpha\beta}\frac{\p X^{\alpha}}{\p x^a} \frac{\p X^{\beta}}{\p x^b}=\Omega^2 g_{ab}
\ee
for ($N+1)$ unknown functions $(X^1, \dots, X^N, \Omega)$ of 
$(x^1, \dots, x^n)$. For the system to admit solutions the number of equation
should equal the number of unknowns, which gives
$N=n(n+1)/2-1$. In the work of  \cite{jacobowitz} this numerology is made precise in the real analytic setup. If $N<n(n+1)/2-1$ then
the system (\ref{system_conf1}) is overdetermined, and in general there will be no solutions. 
If $N\leq n(n+1)/2-1$ is the smallest integer such that
a pseudo--Riemannian manifold $(M, g)$ admits a conformal
embedding in $\R^N$, then its 
{\em conformal embedding class}\footnote{It is clear from this defintion that a conformal embedding class of
any metric $g$ is not greater than its isometric embedding class. The two classes can of course be equal. For
assume that the conformal embedding class of $g$ is $k$. Therefore there exists an $\Omega:M\rightarrow\R^+$ such that
the isometric embedding class of $\hat{g}=\Omega^2g$ is $k$. But the conformal embedding class of $\hat{g}$ is also
$k$. } is $N-n$. 

For a four--dimensional space--time the local conformal embedding is always possible in dimension $r+s=9$ or less. In this paper we shall study local and 
global conformal embeddings of class 1.
Thus we will be interested in embeddings of  vacuum Lorentzian four-manifolds 
$(M, g)$ with a rescaling $\hat{g}=\Omega^2g$ which has isometric 
embedding class 1.

In what follows we shall need the formulae for conformal rescaling in  four dimensions. If $\hat{g}_{ab}=\Omega^2g_{ab}$, and
$\Upsilon_a=\Omega^{-1}\nabla_a\Omega$ these are:
\begin{subequations}
\begin{align}
 \hat{C}_{abc}^{\;\;\;\;\;d}&=C_{abc}^{\;\;\;\;\;d}, \label{c1}\\
 \hat{R}_{ab}&=R_{ab}+2\nabla_a\Upsilon_b-2\Upsilon_a\Upsilon_b+g_{ab}(\nabla_c\Upsilon^c+2\Upsilon_c\Upsilon^c),\label{c2}\\
 \hat{R}&=\Omega^{-2}(R+6\frac{\Box\Omega}{\Omega}),\label{c3}
\end{align}
\end{subequations}
where $\Box$ is the wave operator of $g$, the tensors
$R, R_{ab}, C_{abc}^{\;\;\;\;\;d}$ are respectively the Ricci scalar, the Ricci tensor and the Weyl tensors of $g$,  and the hatted objects correspond to $\hat{g}$. 

Now suppose that the unit normal to $M$ in $\mathbb{R}^5$ is $N_\alpha$ and 
\[\eta^{\alpha\beta}N_\alpha N_\beta=\epsilon=\pm 1,\]
so that the extra dimension will be time-like or space-like according as $\epsilon=1$ or $-1$. There is a projection operator
\[\Pi_\alpha^{\,\beta}=\delta_\alpha^{\,\beta}-\epsilon N_\alpha N^\beta\]
which, with index lowered, as $\Pi_{\alpha\beta}$, pulls back to the metric $\hat{g}_{ab}$ on $M$.

The 5-dimensional covariant derivative of the normal defines the second fundamental form as a tensor on $M$:
\[\hat{K}_{\alpha\beta}:=\Pi_\alpha^{\,\lambda}\Pi_\beta^{\,\mu}\partial_\lambda N_\mu\rightarrow \hat{K}_{ab},\]
where $\partial_\alpha$ is the flat 5-dimensional Levi-Civita covariant derivative and the hat is a reminder that it is $\hat{g}$ that has isometric embedding class 1, and we shall want expressions relating rather to $g$.

As a consequence of the embedding into flat space, one obtains the Gauss equation
\be\label{em1}
\hat{R}_{abcd}=\epsilon(\hat{K}_{ad}\hat{K}_{bc}-\hat{K}_{ac}\hat{K}_{bd}),
\ee
and the Codazzi equation
\be\label{em2}
\hat{\nabla}_{[a}\hat{K}_{b]c}=0,
\ee
by commuting 5-dimensional derivatives and projecting into $M$. The theory for this can be found in chapter 37 of \cite{exsol}, but one way to derive these equations is as follows: in $\mathbb{R}^5$ there are pseudo-Cartesian coordinates $X^\alpha$ satisfying
\[\partial_\alpha X^\beta=\delta_\alpha^\beta, \quad
\mbox{or equivalently} \quad
\partial_\alpha X_\beta=\eta_{\alpha\beta}.\] 
We obtain a co-vector field $\hat{X}_a$ on $M$ by projecting $X_\alpha$ into $M$ and a scalar on $M$ as $\hat{Y}=N^\alpha X_\alpha$; then by projecting the defining equation for 
$X_\alpha$ we obtain the system entirely in $M$:
\begin{subequations}
\begin{align}
\hat{\nabla}_a\hat{X}_b&=\hat{g}_{ab}-\epsilon\hat{Y}\hat{K}_{ab}\label{e1}\\
\hat{\nabla}_a\hat{Y}&=\hat{K}_{ab}\hat{X}^b\label{e2}
\end{align}
\end{subequations}
and this system must have a 5-dimensional vector space of solutions. Commute derivatives on (\ref{e1}) and use (\ref{e1}) and (\ref{e2}) to obtain
\[\hat{R}_{abc}^{\;\;\;\;\;d}\hat{X}_d=\epsilon \hat{Y}\hat{\nabla}_{[a}\hat{K}_{b]c}+\epsilon \hat{K}_{d[a}\hat{K}_{b]c}\hat{X}^d,\]
which has to hold for the 5-dimensional vector space of $(\hat{X}_a,\hat{Y})$ and this is only possible if (\ref{em1}) and (\ref{em2}) hold. Commuting derivatives on (\ref{e2}) gives nothing new. Conversely, if (\ref{em1}) and (\ref{em2}) hold then one can solve (\ref{e1}) 
and (\ref{e2}) to find coordinates for the flat embedding. In the next Proposition we shall use the spinor decomposition of the Weyl tensor
$C_{abcd}=\psi_{ABCD}\epsilon_{A'B'}\epsilon_{C'D'}+\overline{\psi}_{A'B'C'D'}\epsilon_{AB}\epsilon_{CD}$
\begin{prop}
\label{prop0}
Let $\sigma_{ab}=\sigma_{ABA'B'}$ be a symmetric trace--free tensor on a Ricci--flat Lorentzian manifold $(M, g)$ which satisfies
\be\label{em3a}\nabla_{A'(A}\sigma_{BC)B'}^{\;\;\;\;\;\;\;\;A'}=0,\ee
and
\be\label{em4}
\sigma_{(AB}^{\;\;\;\;\;\;C'D'}\sigma_{CD)C'D'}=-2\epsilon\psi_{ABCD} \quad \mbox{where}\quad \epsilon=\pm 1.
\ee
\begin{itemize}
\item
The conditions (\ref{em3a}) and (\ref{em4}) are necessary and sufficient for $(M, g)$ to admit a local conformal embedding  $\iota:M \rightarrow \R^{r, s}$
where $(r, s)$ equals $(1, 4)$ or $(2, 3)$, such that $\hat{g}=\Omega^2 g=\iota^*(\eta)$, and 
the trace--free part of the 2nd fundamental form of the isometric embedding of $(M, \hat{g})$ into $\R^{r, s}$ with $r+s=5$
is given by $\Omega\sigma_{ab}$.
\item Given a solution $\sigma_{ab}$ to (\ref{em3a}, \ref{em4}), there exists a six--dimensional space of pairs $(\Omega, \hat{K})$ giving the conformal  embeddings of $(M, g)$ with the conformal factor $\Omega$, and the mean curvature $\hat{K}=\mbox{trace}(\hat{K}_{ab})$.
\item If $\sigma_{ABA'B'}$ is fixed by an isometry of $(M,g)$ then one may choose $(\Omega,\hat{K})$ also to be fixed by the isometry.
\end{itemize}
\end{prop}
{\bf Proof.}
We want to investigate the system (\ref{em1})-(\ref{em2}). We first decompose $\hat{K}_{ab}$ into its trace-free and trace parts as
\[\hat{K}_{ab}=\hat{\sigma}_{ABA'B'}+\frac14\hat{K}\hat{g}_{ab},\]
and then the trace-free part of (\ref{em2}) is
\be\label{em3}\hat{\nabla}_{A'(A}\hat{\sigma}_{BC)B'}^{\;\;\;\;\;\;\;\;A'}=0.\ee
This equation is conformally invariant: if $\sigma_{ABA'B'}$ has conformal weight 1 so that
\[\hat{\sigma}_{ABA'B'}=\Omega\sigma_{ABA'B'}\mbox{ or equivalently  }\hat{\sigma}_{ABA'}^{\;\;\;\;\;\;\;\;B'}=\sigma_{ABA'}^{\;\;\;\;\;\;\;\;B'}\]
then (\ref{em3}) is preserved. We will write it without hats for future reference as (\ref{em3a})
so that a necessary condition for $(M,g)$ to have conformally embedding class 1 is that there exists a real solution of (\ref{em3a}). 

We note a couple of consequences of (\ref{em3a}). We may decompose the derivative of $\sigma_{ABA'B'}$ into irreducible parts as
\[\nabla_A^{\;\;A'}\sigma_{BC}^{\;\;\;\;\;\;B'C'}=\phi_{ABC}^{\;\;\;\;\;\;\;\;\;A'B'C'}+\epsilon^{A'(B'}\phi_{ABC}^{\;\;\;\;\;\;\;\;\;C')}+\epsilon_{A(B}\overline{\phi}_{C)}^{\;\;A'B'C'}+\epsilon_{A(B}\epsilon^{A'(B'}\phi_{C)}^{\;\;\;C')}, \]
where $\phi_{ABCA'B'C'}$ and $\phi_{ABCC'}$ are symmetric in all indices, and $\phi_{ABC}^{\;\;\;\;\;\;\;\;\;A'B'C'}$ and $\phi_C^{\;\;C'}$ are real. Then the field equation (\ref{em3a}) entails $\phi_{ABCC'}=0$, so that
\begin{equation}\label{em3d}
 \nabla_A^{\;\;A'}\sigma_{BC}^{\;\;\;\;\;\;B'C'}=\phi_{ABC}^{\;\;\;\;\;\;\;\;\;A'B'C'}+\epsilon_{A(B}\epsilon^{A'(B'}\phi_{C)}^{\;\;\;C')}, 
\end{equation}
and by taking the trace on $AB$ and $A'B'$ 
we find
\begin{equation}\label{em3e}
 \phi_a=\phi_{AA'}=\frac{4}{9}\nabla^b\sigma_{ab}.
\end{equation}
If we take the trace of (\ref{em3d}) just on $A'B'$ we find
\[\nabla_{AA'}\sigma_{BC}^{\;\;\;\;\;\;A'C'}=\frac{3}{2}\epsilon_{A(B}\phi_{C)C'},  \]
which can be expressed in terms of tensors, with the aid of (\ref{em3e}) as
\be\label{em3c}
\nabla_{[a}\sigma_{b]c}=\frac13g_{c[a}\nabla^d\sigma_{b]d},\ee
which is useful below.

The trace-free part of the Gauss equation, (\ref{em1}), now gives (\ref{em4})
where both sides have conformal weight zero. Note also that identically
\begin{equation}
\label{id1}
\sigma_{A(B}^{\;\;\;\;\;\;C'D'}\sigma_{CD)C'D'}=\sigma_{(AB}^{\;\;\;\;\;\;C'D'}\sigma_{CD)C'D'}.
\end{equation}

When $(M,g)$ is vacuum we can exploit the vacuum Bianchi identity:
\[\nabla_{A'}^{\;\;A}\psi_{ABCD}=0,\]
to obtain another restriction on $\sigma_{ABA'B'}$ namely (using (\ref{id1}))
\[\nabla_{A'}^{\;\;A}(\sigma_{A(B}^{\;\;\;\;\;\;\;\;\;C'D'}\sigma_{CD)C'D'})=0, \]
which with the aid of (\ref{em3d}) gives
\begin{equation}\label{em3f}
\sigma_{A(B}^{\;\;\;\;\;\;\;\;\;C'D'}\phi^A_{\;\;CD)A'C'D'}=\frac52\phi_{(B}^{\;\;D'}\sigma_{CD)A'D'}, 
\end{equation}
an identity that we shall need below.

To summarise: (\ref{em3a}) and (\ref{em4}) are conformally-invariant necessary conditions for the metric $g$ to have conformal embedding class 1. Given solutions of these, we still have to find $\Omega$ and $\hat{K}$ 
and we have equations from the traces of (\ref{em1}) and (\ref{em2}) available to impose on these.




\medskip
We shall now prove the second part of Proposition \ref{prop0}, by finding and solving equations for $(\Omega,\hat{K})$.
We shall prolong these equations to a connection on a rank $7$ vector bundle over $M$, and show that 
the parallel sections of this connection,
subject to one  algebraic constraint, correspond to solutions of our system.

From the trace of (\ref{em2}) we obtain
\be\label{em5}\nabla_a\hat{K}=\frac43\Omega^{-1}(\nabla_b\sigma_a^{\;\;b}+3\Upsilon_b\sigma_a^{\;\;b}),\ee
and from the trace of (\ref{em1}) we obtain
\be\label{em6}\hat{R}_{ab}=\epsilon(\sigma_{ac}\sigma_b^{\;c}-\frac12\Omega\hat{K}\sigma_{ab}-\frac{3}{16}\Omega^2\hat{K}^2g_{ab}),\ee
where the right-hand-side is computed entirely using $g_{ab}$ and $\nabla_a$, and from (\ref{c3})
\[\hat{R}_{ab}=R_{ab}+2\nabla_a\Upsilon_b+g_{ab}\nabla_c\Upsilon^c-2\Upsilon_a\Upsilon_b+2g_{ab}\Upsilon_c\Upsilon^c,\]
or, introducing $\Theta=\Omega^{-1}$,
\be\label{em7}  \hat{R}_{ab}=R_{ab}-2\Omega\nabla_a\nabla_b\Theta+g_{ab}(-\Omega\Box\Theta+3\Omega^2|\nabla\Theta|^2).\ee

To check the integrability conditions for $\hat{K}_a$ we first calculate
\[\nabla_{[a}\nabla^c\sigma_{b]c}= (\nabla_{[a}\nabla^c-\nabla^c\nabla_{[a})\sigma_{b]c}+\nabla^c(\nabla_{[a}\sigma_{b]c}). \]
The curvature terms vanish if $(M,g)$ has vanishing trace-free Ricci tensor, and the third term is expressible with the aid of (\ref{em3a}) as
\[\nabla^c(\frac13\nabla^d\sigma_{d[a}g_{b]c]})=-\frac13\nabla_{[b}\nabla^c\sigma_{a]c},\]
so this is zero and the 1-form $\nabla^b\sigma_{ab}$ is closed. Now we turn to (\ref{em5}) written as 
\[\frac34\nabla_a\hat{K}=\Theta(\nabla_b\sigma_a^{\;\;b}+3\Upsilon_b\sigma_a^{\;\;b})=\Theta\nabla^c\sigma_{ac}-3\Theta^c\sigma_{ac},\]
and take a curl:
\[\frac34\nabla_{[a}\nabla_{b]}\hat{K}=\Theta_{[a}\nabla^c\sigma_{b]c}-3(\nabla_{[a}\Theta^c)\sigma_{b]c}-3\Theta^c\nabla_{[a}\sigma_{b]c}.\]
The second derivatives of $\Theta$ can be eliminated with the aid of (\ref{em7}), provided $(M,g)$ has vanishing trace-free Ricci tensor, and the first derivatives cancel with the aid of (\ref{em3c}). Thus 
integrability for $\hat{K}$ follows, at least when $(M,g)$ is Einstein.\\

To check integrability for $\Theta$ we rewrite (\ref{em7}) and (\ref{em6}) together as
\begin{equation}\label{em8}\nabla_a\nabla_b\Theta=-\frac{\epsilon\Theta}{2}(\sigma_{ac}\sigma_b^{\;\;\;c}-\frac12\Omega\hat{K}\sigma_{ab})+
g_{ab}\left(\frac12\Omega|\nabla\Theta|^2+\frac{\epsilon\Theta}{12}\sigma_{cd}\sigma^{cd}+\frac{\epsilon}{32}\Omega\hat{K}^2\right), \end{equation}
and there will be integrability conditions for this. To see what they are we prolong to obtain a linear system for $(\hat{K},\Theta,\Theta_a,H)$ where $H$ is to be defined. We have at once
\begin{eqnarray}
 \nabla_a\hat{K}&=&\frac43\Theta\nabla^c\sigma_{ac}-4\Theta^c\sigma_{ac},\label{sys1}\\
 \nabla_a\Theta&=&\Theta_a,\label{sys2}\\
 \nabla_a\Theta_b&=&-\frac12\epsilon\Theta\sigma_{ac}\sigma_b^{\;\;\;c}+\frac14\epsilon\hat{K}\sigma_{ab}+g_{ab}(\frac{1}{12}\epsilon\Theta\sigma_{cd}\sigma^{cd}+H),\label{sys3} 
\end{eqnarray}
with
\[H=\frac12\Theta^{-1}|\nabla\Theta|^2+\frac{1}{32}\epsilon\Theta^{-1}\hat{K}^2,\]
making use of (\ref{em5}) and (\ref{em8}). Note that $H$ is fixed by the vanishing of the quadratic
\begin{equation}\label{em10}
Q:=H\Theta-\frac{1}{2}g^{ab}\Theta_a\Theta_b-\frac{\epsilon}{32}\hat{K}^2,
 \end{equation}
We need an equation for $\nabla_aH$, and using other equations in the system we find this to be
\begin{equation}\label{sys4}
 \nabla_aH=-\frac12\epsilon\sigma_{ac}\sigma_b^{\;\;\;c}\Theta^b+\frac{1}{12}\epsilon\Theta_a\sigma_{cd}\sigma^{cd}+\frac{1}{12}\epsilon\hat{K}\nabla^b\sigma_{ab}.
\end{equation}
We need to calculate the curvature of the connection defined by (\ref{sys1})-(\ref{sys4}). The commutator of derivatives on (\ref{sys1}) has been seen to be zero; on (\ref{sys2}) it vanishes by virtue of (\ref{sys3}); on (\ref{sys3}) and (\ref{sys4}) we make use of the identities (\ref{em3d})-(\ref{em3f}) and after a straightforward but lengthy calculation find that they too vanish. Thus, given $(\hat{K},\Theta,\Theta_a)$ at an initial point $p$, we obtain $H$ from the vanishing of $Q$ in (\ref{em10}) and obtain $(\hat{K},\Theta)$ in a neighbourhood of $p$ by line integrals.

For the last part, given an isometry $\varphi$ of $(M,g)$, if it preserves $\sigma_{ABA'B'}$ then it preserves the coefficients of the connection defined by (\ref{sys1})-(\ref{sys4}). Thus, if we choose data preserved by the isometry (in the sense that ${\mathcal{L}}_\varphi\hat{K}, {\mathcal{L}}_\varphi\Theta$ and ${\mathcal{L}}_\varphi\Theta_a$ vanish initially) then necessarily they vanish everywhere.
\koniec
\subsection{Algebraic obstructions for conformal class 1} 
\label{subalg}
There are some algebraic conditions the Weyl tensor needs
to satisfy in order that solutions to (\ref{em4}) exist. In this section we shall find all these conditions - they will
be neccessary and sufficient for the existence of $\sigma_{ab}$ such that (\ref{em4}) holds,  but only necessary for
the existence of a class 1 conformal embedding,  as there may be other obstructions coming from the differential
condition (\ref{em3a}). To make the obstructions applicable to Lorentzian as well  as Riemannian signatures of $g$ we shall
(only in this section) consider $(M, g)$ to be a holomorphic Riemannian manifold,  where the anti--self--dual Weyl spinor
$\psi_{ABCD}$ and the  self--dual Weyl spinor $\psi_{A'B'C'D'}$ are independent. In this case the trace--free part of
the Gauss equation (\ref{em1}) gives the system
\be
\label{em4c}
\sigma_{(AB}^{\;\;\;\;\;\;C'D'}\sigma_{CD)C'D'}=-2\epsilon\psi_{ABCD}, \quad
\sigma_{(A'B'}^{\;\;\;\;\;\;CD}\sigma_{C'D')CD}=-2\epsilon\psi_{A'B'C'D'}.  
\ee
There are four algebraic invariants (see \cite{pr}) of the Weyl spinors:
\begin{eqnarray}
\label{IJ}
I&=&\psi_{ABCD}\psi^{ABCD}, \quad J={\psi_{AB}}^{CD}{\psi_{CD}}^{EF}{\psi_{EF}}^{AB},\\
I'&=&\psi_{A'B'C'D'}\psi^{A'B'C'D'}, \quad J'={\psi_{A'B'}}^{C'D'}{\psi_{C'D'}}^{E'F'}{\psi_{E'F'}}^{A'B'},\nonumber
\end{eqnarray}
which are in general independent. We verify by explicit calculation that if $\psi_{ABCD}$ and $\psi_{A'B'C'D'}$
arise from $\sigma_{ABA'B'}$ by (\ref{em4c}), then $I=I'$, and $J=J'$. In fact these conditions
are (at the algebraic level) also sufficient for the existence of $\sigma_{ABA'B'}$. To see this,  note
that  the system (\ref{em4c}) does not determine $\sigma_{ABA'B'}\sigma^{ABA'B'}$, and so it consists
of ten equations for eight unknowns. By computing the wedge product of differentials
of equations (\ref{em4c}) we show that any nine out of ten equations are algebraically dependent, but any
eight of ten equations are independent. Thus we can pick eight equations, and solve them for
eight components of $\sigma_{ABA'B'}$ in terms of the components $\psi_{ABCD}$ and $\psi_{A'B'C'D'}$. Substituting the resulting expressions in the remaining two equations yields at most two algebraic conditions
on the Weyl spinors. But we have already found two such conditions, so we have established
\begin{prop}
\label{propcomplex}
Let $I, J, I', J'$ be the invariants (\ref{IJ}) of the Weyl spinors of $(M, g)$. The conditions 
\be
\label{IeqJ}
I=I', \quad J=J'
\ee
are necessary and sufficient for the existence of $\sigma_{ABA'B'}$ such that (\ref{em4c}) holds. These conditions
are necessary for the existence of the class one conformal embedding of $(M, g)$ in five dimensions.
\end{prop}
By imposing Riemannian and Lorentzian reality conditions on Proposition \ref{propcomplex} we deduce the following
\begin{col}
A Riemannian four--manifold with anti--self--dual Weyl tensor admits a class one conformal embedding if and only if
it is conformally flat.
\end{col}
{\bf Proof.} The anti--self--duality of the Weyl tensor is equivalent to the spinor condition
$\psi_{A'B'C'D'}=0$. Therefore $I'=0$. However  in the Riemannian signature $I=0$ if and only if $\psi_{ABCD}$
vanishes. Thus conformal flatness is necessary and sufficient for the existence of $\sigma_{ABA'B'}$ in this case.
\koniec
\begin{prop}
\label{prop_real}
Let $(M, g)$ be a Lorentzian four--manifold. The necessary and sufficient conditions for the existence
of $\sigma_{ABA'B'}$ such that (\ref{em4}) holds are that $I$ and $J$ are real.  Thus the reality of $I$ and $J$ is 
necessary for the existence of class one  conformal embeddings.
\end{prop}
{\bf Proof.} The `necessary' part follows directly from Proposition \ref{propcomplex}, as in Lorentzian
signature $I=I'$ and $J=J'$. 
To establish sufficiency we need to show that given real $(I, J)$ there
exists a solution to (\ref{em4}) which is also real in a sutiable sense.
The analysis is quite tedious, and the details depend on the algebraic
type of the Weyl spinor. We give it in Appendix A.
\koniec
These results can be used to rule out the existence of class one conformal embeddings for several 
known solutions to Einstein equations.
\begin{col}
\label{kercol}
The Lorentzian Kerr metric, the Riemannian anti-self-dual Taub--NUT metric, and the Riemannian Fubini--Study metric on $\CP^2$
do not admit local class one conformal embeddings\footnote{Both Taub--NUT and $\CP^2$ can be conformally 
embedded in $\R^7$, \cite{DT191}.}.
\end{col}
{\bf Proof.} In the case of Kerr we find that $I$ is not real. Both Fubini--Study and ASD Taub--NUT are ASD and not conformally flat.
\koniec
\subsection{The necessary conditions applied to the Schwarzschild metric}
What follows will work in any static, spherically symmetric metric but we restrict to the Schwarzschild solution for simplicity. We will work with the Kruskal form of the metric in order to embed the largest possible piece of 
the Schwarzschild metric, so suppose this form is
\be\label{sc1} g=2F^2(r)dudv-4r^2\frac{d\zeta d\overline{\zeta}}{P^2},\ee
where $P=1+\zeta\overline{\zeta}$, $\zeta= e^{i\phi}\tan{\theta/2}$ and the null coordinates $u,v$ are connected to the usual Schwarzschild $t,r$ by
\[u=-e^{(r-t)/4m}(\frac{r}{2m}-1)^{1/2},\;\;v=e^{(r+t)/4m}(\frac{r}{2m}-1)^{1/2},\]
and finally 
\[F^2=16\frac{m^3}{r}e^{-r/2m}.\]

We first obtain the following:
\begin{theo}
\label{theop}
If the conformal embedding of Proposition \ref{prop0} is global on at least one sphere of symmetry of the metric
(\ref{sc1}) then $\sigma_{ABA'B'}$ is necessarily spherically symmetric, and the embedding can be chosen to be spherically symmetric.
\end{theo}
We shall prove this in two steps. First we shall show that the assumptions imply the spherical symmetry of $\sigma_{ab}$, and
then we shall deduce from Proposition \ref{prop0} that the conformal factor and the mean curvature also need to be spherically symmetric.

 We shall calculate in the GHP formalism \cite{ghp}, which we summarise in Appendix  B. 
We want to expand (\ref{em3}) and (\ref{em4}) in this formalism. We begin by expanding $\sigma_{ABA'B'}$:
\[\sigma_{ABA'B'}=Xo_Ao_Bo_{A'}o_{B'}+Uo_Ao_Bo_{(A'}\iota_{B')}+To_Ao_B\iota_{A'}\iota_{B'}\]
\[+\overline{U}o_{(B}\iota_{B)}o_{A'}o_{B'}+2Yo_{(A}\iota_{B)}o_{(A'}\iota_{B')}+\overline{U}'o_{(A}\iota_{B)}\iota_{A'}\iota_{B'}\]
\[+\overline{T}\iota_A\iota_B o_{A'} o_{B'}+U'\iota_A\iota_Bo_{(A'}\iota_{B')}+X'\iota_A\iota_B\iota_{A'}\iota_{B'}.\]
Here $X,X'$ and $Y$ are real, with GHP weights $(-2,-2),(2,2),(0,0)$ respectively, and so  
have zero spin weight while $U,U',T$ have nonzero spin weight, and our first task is to show that the last three 
vanish if the embedding is global on at least one of the spheres of symmetry. We write out (\ref{em3}) at length, obtaining a system of four equations (see Appendix B for notation):
\begin{subequations}
\begin{align}
\et'X+\frac12\th'U-\rho'U&=0\label{g3}\\
\frac12\et'U+\th'T-\rho'T&=0\label{g4}\\
-\th X-\frac12\et U+\et'\overline{U}+\th'Y-\rho X-\rho'Y&=0\label{g5}\\
-\frac12\th U+\et'Y+\th'\overline{U}'-\et T&=0\label{g6}
\end{align}
\end{subequations}
together with their primes (here $T'=\overline{T}$):
\begin{subequations}
\begin{align}
\et X'+\frac12\th U'-\rho U'&=0\label{g7}\\
\frac12\et U'+\th T'-\rho T'&=0\label{g8}\\
-\th' X'-\frac12\et'U'+\et\overline{U}'+\th Y-\rho' X'-\rho Y&=0\label{g9}\\
-\frac12\th' U'+\et Y+\th\overline{U}-\et'T'&=0\label{g10}.
\end{align}
\end{subequations}
To set about solving these equations, we use the fact that $\et$ is onto from spin-weight 0 to spin-weight 1 and in fact from $s$ to $s+1$ with $s\geq 0$, 
with the corresponding statement for $\et'$. Thus we can introduce potentials 
$W,W'$ and $Q$ with
\[U=\et'W,\;\;\;\;U'=\et W',\;\;\;\;T=\et'^2Q,\]
where conventionally $Q'=\overline{Q}$. Equation (\ref{g3}) becomes
\[\et'X=-\frac12\th'\et'W+\rho'W=-\frac12\et'\th'W+\frac12\rho'W=\et'(-\frac12\th'W+\frac12\rho'W),\]
making use of the commutators and spherical symmetry of $\rho'$, so that
\be\label{x1}\et'(X+\frac12\th'W-\frac12\rho'W)=0.\ee
On a sphere of symmetry one has complete information about the angular dependence of smooth functions in the kernel of powers of $\et$ or $\et'$, and indeed of eigenfunctions and eigenvalues of the Laplacians $\et\et'$ 
or $\et'\et$. For spin-weight zero functions, the eigenfunction equation of the Laplacian is
\[\et\et'f=\et'\et f=\ell(\ell+1)(\rho\rho'+\psi_2)f,\]
for non-negative integer $\ell$, and the kernel of $\et^k$ is spanned by these eigenfunctions with $0\leq \ell<k$. 

In particular we can deduce from (\ref{x1}) that
\be\label{g11}X=-\frac12\th'W+\frac12\rho'W+X_0\ee
where $X_0$ is constant in the angles. Similarly from (\ref{g7})
\be\label{g12}X'=-\frac12\th W'+\frac12\rho W'+X'_0,\ee
with $X'_0$ independent of angles.

From (\ref{g4}) we obtain by similar manipulations
\[\et'^2(\frac12W+\th'Q+\rho'Q)=0,\]
so that 
\[W=-2(\th'Q+\rho'Q)+W_0\]
where $W_0$ is a combination of $\ell=0$ and $1$ spherical harmonics. Since $W$ is undefined up to additive constant (in the angles) we can suppose that $W_0$ is purely $\ell=1$. Then from (\ref{g8})
\[W'=-2(\th\overline{Q}+\rho\overline{Q})+W_0'\]
with $W_0'$ purely $\ell=1$. From its definition, $Q$ may be assumed to contain only terms of $\ell\geq 2$ and so $W$ and $W'$ contain only terms of $\ell\geq 1$.

From the imaginary part of (\ref{g5})
\[0=\et U-\et'\overline{U}=\et'\et(W-\overline{W}),\]
but $\et'\et$ here is (half) the Laplacian which has trivial kernel on $S^2$ and we deduce that $W$ is real, as from (\ref{g9}) is $W'$.

From (\ref{g6}) by now familiar methods we deduce
\be\label{g13}Y=\frac12(\th+\rho)W-(\th'+\rho')W'+(\et\et'-2(\rho\rho'+\psi_2))Q+Y_0\ee
with $Y_0$ independent of angles, and from (\ref{g10})
\be\label{g14}Y=\frac12(\th'+\rho')W'-(\th+\rho)W+(\et'\et-2(\rho\rho'+\psi_2))\overline{Q}+\widetilde{Y}_0,\ee
where again $\widetilde{Y}_0$ is independent of angles, and in both of these we've used reality of $W$ and $W'$. Recall that $Y$ is real, but the imaginary part of $Y$ from (\ref{g13}) is
\[(Y_0-\overline{Y}_0)+(\et\et'-2(\rho\rho'+\psi_2))(Q-\overline{Q}),\]
which must therefore vanish. However the first bracket contains only $\ell=0$ terms while the rest has only $\ell\geq 2$, so these must vanish separately. The operator acting on $(Q-\overline{Q})$ has trivial kernel 
on functions with $\ell\geq 2$ so we conclude that $Q$ and $Y_0$ are real. 

By comparing (\ref{g13}) and (\ref{g14}) with $Q,Y_0,\widetilde{Y}_0$ all real, we also conclude that $Y_0=\widetilde{Y}_0$ and 
\be\label{g15}(\th'+\rho')W'=(\th+\rho)W.\ee
Using this in (\ref{g5}) and looking only at $\ell=1$ terms we calculate
\[6\psi_2W_0=0\]
and since $m\neq 0$ we conclude $W_0=0$, when similarly from (\ref{g9}) $W'_0=0$.

Summarising we have
\[X=(\th'-\rho')(\th'+\rho')Q+X_0,\;\;X'=(\th-\rho)(\th+\rho)Q+X_0',\]
\[Y=(\th+\rho)(\th'+\rho')Q+(\et\et'-2(\rho\rho'+\psi_2))Q+Y_0\]
\[U=-2\et'(\th'+\rho')Q,\;\;U'=-2\et(\th+\rho)Q,\;\;T=\et'^2Q.\]

We substitute into (\ref{g5}) and (\ref{g9}) and keep only terms in $Q$ (i.e. with $\ell\geq 2$) to obtain
\[-6\psi_2(\th'+\rho')Q=0= -6\psi_2(\th+\rho)Q.\]
Since $m\neq 0$ these force \[(\th'+\rho')Q=0=(\th+\rho)Q\] when many things follow:
\begin{itemize}
 \item $U=U'=0$ and the $Q$ contributions to $X$ and $X'$ vanish;
 \item one can integrate to find $Q=rq(\theta,\phi)$ for some $q$ with $\ell\geq 2$;
 \item now $T=\frac{1}{r}\et_0'^2q$ where $\et_0$ is $\et$ on the unit sphere (which is independent of $r$) and 
 \[Y=Y_0+\frac{1}{r}(\et_0\et'_0+2)q,\]
 \item what is left of (\ref{g5}) and (\ref{g9}) becomes
 \[(\th+\rho)X-(\th'-\rho')Y=0=(\th'+\rho')X'-(\th-\rho)Y.\]
\end{itemize}
Note that, if $X=0=X'$ we still have
\[(\th'-\rho')Y=0=(\th-\rho)Y,\]
which solve as
\[Y=\frac{y(\theta,\phi)}{r},\]
for some $y)\theta,\phi)$. The $q$-dependent part of $Y$ automatically has this form but we learn that $Y_0$ does too.

At this point we turn to the algebraic conditions (\ref{em4}):
\[\sigma_{(AB}^{\;\;\;\;\;\;C'D'}\sigma_{CD)C'D'}=-2\epsilon\psi_{ABCD}.\]
With the chosen $\sigma_{ABA'B'}$ and the type D Weyl spinor of Schwarzschild, and taking account of $U=0=U'$ all we have left is
\[XT=0=X'T,\;\;XX'-Y^2+|T|^2=-6\epsilon\psi_2.\]
One possibility is evidently $T=0$ in which case $Q=0$ and $\sigma_{ABA'B'}$ is spherically-symmetric with
\[(\th+\rho)X_0-(\th'-\rho')Y_0=0=(\th'+\rho')X'_0-(\th-\rho)Y_0.\]
Are there other possibilities? If $T\neq 0$ then $X=X'=0$ and 
\[(Y_0+\frac{A}{r})^2-\frac{|B|^2}{r^2}= 6\epsilon\psi_2=-\frac{6\epsilon m}{r^3},\]
writing $A,B$ for $(\et_0\et'_0+2)q,\et_0'^2q$ respectively. However, in this case we know the $r$-dependence of $Y_0$ and the left-hand-side of this expression is proportional to $r^{-2}$ while the right-hand-side 
is proportional to $r^{-3}$ -- a contradiction. Thus $T=0$ and $\sigma_{ABA'B'}$ is spherically symmetric, of the form:

\be\label{g16}\sigma_{ABA'B'}=Xo_Ao_Bo_{A'}o_{B'} +2Yo_{(A}\iota_{B)}o_{(A'}\iota_{B')} +X'\iota_A\iota_B\iota_{A'}\iota_{B'},\ee
with $X,X',Y$ functions only of $u,v$ and subject to
\be\label{g17}(\th+\rho)X-(\th'-\rho')Y=0=(\th'+\rho')X'-(\th-\rho)Y,\ee
and 
\be\label{g18}XX'-Y^2=-6\epsilon\psi_2=\frac{6\epsilon m}{r^3}.\ee
In coordinates (\ref{g17})
becomes
\be\label{g19}\frac{r^2}{F^2}\left(\frac{F^2X}{r}\right)_u-(rY)_v=0=\frac{r^2}{F^2}\left(\frac{F^2X'}{r}\right)_v-(rY)_u.\ee

\medskip

\medskip

With $\sigma_{ab}$ spherically symmetric we know from Proposition \ref{prop0} that  $\Omega$ and $\hat{K}$ can be chosen also to be spherically symmetric.
%
\koniec
It is a simple application of the Cauchy-Kowalewski Theorem to see that analytic solutions of the system (\ref{g18})-(\ref{g19}) depend on two free analytic functions of one variable. To see this, note that $r$ is analytic in $uv$, so that (\ref{g19}) can be written
\[X_u=F_1(u,v,X,Y,Y_v),\;\;Y_u=F_2(u,v,Y,X',X'_v),\]
while the $u$-derivative of (\ref{g18}) can be solved for $X'_u$:
\[X'_u=F_3(u,v,X,Y,X',Y_v,X'_v),\]
and the functions $F_i$ are analytic in all arguments away from $r=0$ and $X=0$. We may choose analytic data $X(0,v),Y(0,v)$ with $X(0,v)\neq 0$ on an interval, say $I$, in $v$ on the line $u=0$. Then (\ref{g18}) can be solved for $X'(0,v)$ on $I$ and the Cauchy-Kowalewski Theorem provides an analytic solution on a neighbourhood of $I$. The equation (\ref{g18}) is preserved by the system. With an appropriate 
choice of $I$, this will give a solution covering the bifurcation surface at $u=v=0$.

For later use, 
we calculate the divergence:
\[\nabla_b\sigma_a^{\;b}=\ell_a(\th X+\frac12\th'Y-2\rho X-2\rho'Y)+n_a(\th'X'+\frac12\th Y-2\rho'X'-2\rho Y),\]

\subsubsection{Imposing staticity on the embedding}\label{sssSE}
The embedding is not forced to be static but we can impose it. The time-like Killing vector of the Schwarzschild metric (\ref{sc1}) is
\[T:=4m\partial_t=u\partial_u-v\partial_v.\]
This does not preserve the null basis: calculate
\[[T,\ell]=-\ell,\;\;[T,n]=n,\;\;[T,m]=0.\]
Now with
\[\sigma_{ab}=X\ell_a\ell_b+Y(\ell_an_b+n_a\ell_b-\frac12g_{ab})+X'n_an_b\]
obtain
\[\mathcal{L}_T\sigma_{ab}=(T(X)-2X)\ell_a\ell_b+T(Y)(\ell_an_b+n_a\ell_b-\frac12g_{ab})+(T(X')+2X')n_an_b.\]
Set this to zero and solve to see that a static embedding is equivalent to the choices
\begin{equation}\label{ss1}
X=u^2f(r),\;\;Y=y(r),\;\;X'=v^2g(r).
\end{equation}
Impose (\ref{g19}) then
\[-\frac{(r-2m)}{r^3}(ry)'= \left(\frac{(r-2m)^2}{r^2}e^{r/2m}g\right)'= \left(\frac{(r-2m)^2}{r^2}e^{r/2m}f\right)'.\]
This implies
\[f-g=\frac{c_1r^2}{(r-2m)^2}e^{-r/2m},\]
with a constant of integration $c_1$, so for solutions bounded at $r=2m$ we need $f=g$. How does this relate to the algebraic condition?

\medskip

This is
\[(ry)^2=(ruvf)^2-6\epsilon m/r,\]
so eliminate $ry$ and differentiate to get a 1st-order ODE for $f$. For bounded $f$ extending to the horizon (where $uv=0$) this forces $\epsilon=-1$.

From Proposition \ref{prop0} we know we may choose $\Omega$ and $\hat{K}$ to be static and therefore functions only of $r$. Now substituting $\sigma_{ABA'B'}$ and $\hat{K}$ into the system (\ref{e1})-(\ref{e2}), suitably rescaled with $\Omega$ we find the following five linearly independent flat coordinates:
\[u\Omega F,\;v\Omega F,\;r\Omega\sin\theta\cos\phi,\; r\Omega\sin\theta\sin\phi,\;r\Omega\cos\theta,\]
which we recognise as the embedding in section \ref{ekc}.
\subsection{Rigidity of class 1 conformal embeddings}
The isometric embedding is called {\em rigid} if it is unique up to an isometry of the ambient space. It has been
shown by Thomas \cite{thomas} that class 1 isometric embeddings are rigid in a neighbourhood of a point $p\in M$
if they are generic,  i. e. the rank of the 2nd 
fundamental form at $p$ is maximal\footnote{Isometric embeddings of higher co-dimensions need not be rigid even if they are generic - 
see \cite{bryant} for details.}.
We shall use this result to discuss the rigidity of the class 1 conformal embeddings. 

Recall that Proposition \ref{prop0}
splits the construction of such embeddings into two steps 
\begin{itemize}
\item[(A)] Find an isometric embedding of a conformally rescalled metric $\hat{g}=\Omega^2 g$ with  a given trace--free part
of the second fundamental form.
\item[(B)] Reconstruct the mean curvature of the embedding, and the conformal factor.
\end{itemize}
We have shown that once $(A)$ can be achieved, then the space of $(\Omega, \hat{K})$ in $(B)$ is six dimensional. The result of Thomas
above implies that  if the isometric embedding $(A)$ is generic, then it is rigid and therefore depends on $15$  constants  - 
the parameters of the isometry group of $\R^{r, s}$ with $r+s=5$. Thus if the second fundamental form $\hat{K}_{ab}$ has maximal rank,
then the conformal embedding depends on at most $21$ parameters, which is the dimension of the conformal group
$SO(r+1, s+1)$. Indeed it depends on exactly $21$ parameters as  conformal transformations of $\R^{r, s}$ preserve
the conformal class of $\eta$, and so map one conformal embedding into another one, possibly with a different conformal factor.
Therefore our argument shows that there are no more free parameters than one would expect from the conformal motions.
This proves
\begin{prop}
Let $\iota: M\rightarrow \R^{r, s}$ with $r+s=5$ be a local conformal embedding of Proposition \ref{prop0} such that
the rank of the second--fundamental form $\hat{K}_{ab}$ is maximal at some point $p\in M$. Then $\iota$ is rigid in a neighbourhood of $p$
up to conformal transformations of $\R^{r, s}$.
\end{prop}
To this end we note that local conformal embeddings which preserve spherical symmetry are not rigid, as the genericity assumption is not satisfied. The argument below is valid in any dimension. Let $M$ be an $n$--dimensional manifold with a Lorentzian $SO(n-1)$--invariant
metric 
\[
g=V(r)dt^2- W(r)dr^2-r^2\gamma_{S^{n-2}},
\]
where $\gamma_{S^{n-2}}$ is the round metric on $(n-2)$--dimensional sphere, and $V, W$ are arbitrary functions of $r$.  Consider a local conformal embedding $\iota:M\rightarrow
\R^{n, 1}$ given by
\[
\Omega^2 g=dT^2-dX^2-dR^2-R^2 \gamma_{S^{n-2}}, \quad \mbox{where}\quad \Omega^2= R^2/r^2.
\]
The problem of finding $\iota$ readily reduces to an isometric embedding of a surface with a metric
$r^{-2}(V(r)dt^2-W(r)dr^2)$ into a patch in $AdS_3$ with the metric $R^{-2}(dT^2-dX^2-dR^2)$. Isometric
embeddings of surfaces into 3 dimensions (curved or flat) depend, in real analytic category, on arbitrary functions of one variable.
Therefore the conformal embedding $\iota$ is not rigid. An example of a conformal embedding from this class will be disussed
in the next Section.
\section{Conformal embeddings of spherically symmetric metrics}
\label{section1}
In this section we shall construct explicit local conformal embeddings of
spherically symmetric space-times as hypersurfaces in $\R^{4, 1}$.
\begin{prop}
\label{prop1}
A spherically symmetric Lorentizan manifold
$(M, g)$, with
\be
\label{four_m}
g= Vdt^2- V^{-1}dr^2-r^2(d\theta^2+\sin{\theta}^2d\phi^2), \quad\mbox{where}\quad V=V(r)
\ee
can be locally conformally embedded in $\R^{4, 1}$. 
If $V$ has a finite number of simple zeroes at $r_0>r_1>r_2\dots$ then the embedding extends through $r_0$. 
If additionally $g$ is asymptotically flat with $V\rightarrow 1$ as $r\rightarrow \infty$, then the embedding  is
also asymptotically flat and $\Omega\rightarrow 1$ as $r\rightarrow\infty$.
\end{prop}
{\bf Proof.} We shall prove this proposition by reducing the problem to a quadrature, and constructing the embedding explicitly. Consider the conformally flat 5-metric
\be
\label{5-met}
G=\Omega^{-2}(dT^2-dX^2-dR^2-R^2 (d\theta^2+\sin{\theta}^2d\phi^2)).
\ee
We aim to isometrically embed a spherically symmetric Lorentzian manifold
$(M, g)$, with $g$ given by (\ref{four_m})
in $(\R^{4, 1}, G)$. Set $\Omega^{-2}R^2=r^2$. The problem then reduces to finding
an isometric embedding of a two--metric
\[
g_2=
\frac{1}{r^2}(V^{-1}dr^2-Vdt^2)
\]
in a patch of $AdS_3$ with the metric
\[
G_3=\frac{dR^2+dX^2-dT^2}{R^2}.
\]
Setting
\be
\label{eeq1}
T=\sinh{(ta)}f(r), \quad X=\cosh{(ta)}f(r), \quad R= h(r),
\ee
where $a$ is a constant, 
and comparing the coefficients of $dt^2$ gives
\[
f(r)=\frac{h(r)}{ar}\sqrt{V(r)}.
\]
The coefficient of $dr^2$ gives
\be
\label{h_eq}
h=\exp\Big(\int
\frac{V(2V-rV')\pm ar\sqrt{V(4V+4a^2r^2 -(2V-rV')^2)}}{2rV(a^2r^2+V)}dr\Big).
\ee
We have still not made a choice of $a$. Let us assume that $V(r)$ has a finite number
of isolated simple zeroes $r_0>r_1>r_2, \dots\;$. Then the zero at $r=\bar{r}$ of $V$ in the denominator of (\ref{h_eq}) with cancel with a zero of a numerator if
\be
\label{hsa}
a=\pm\frac{1}{2}V'|_{r=\bar{r}},
\ee
which is the surface gravity of (\ref{four_m}) at the Killing horizon $r=\bar{r}$. 

We also claim that the embedding is regular at points where $V+a^2r^2=0$. This can be seen by substituting $V=-a^2r^2$
into (\ref{h_eq}) and leaving $V'$ unspecified. By taking a negative square root the singularity in the denominator
in the integrand then cancels.

If we further assume that the $(M, g)$ is asymptotically flat with 
$V\rightarrow 1, V'\rightarrow 0$ as $r\rightarrow \infty$, then
$h\rightarrow \mbox{const}\cdot r$. We choose the constant of integration to be
$1$ so that the conformal factor $\Omega\rightarrow 1$, and the embedding is asymptotically flat.
\koniec
\section{Global conformal embedding of Schwarzschild}\label{section4}
Rewrite the result of  \S\ref{section1} as
\[
\Omega^{2}\Big(Vdt^2-V^{-1}dr^2-r^2(d\theta^2+\sin^2{\theta}d\phi^2)\Big)
=\iota^{*}(\eta^{\mu\nu}dX_\mu dX_\nu),
\]
where $\eta=\mbox{diag}(1, -1, -1, -1, -1)$, and now $\Omega$ is the pull--back of the conformal factor by $\iota$. 
Consider the Schwarzschild solution which corresponds to 
\[
V=1-\frac{2m}{r}.
\]
In this case
the conformal embedding of \S\ref{section1} is global, that is
the Lorentzian metric $\hat{g}=\Omega^{2}g_{\mbox{schw}}$ conformal
to the Schwarzschild metric is isometrically and globally embedded in
$\R^{4, 1}$. Set 
\[
(X_0, X_1, X_2, X_3, X_4)=(T, X, R\sin{\Theta}\sin{\Phi}, R\sin{\Theta}\cos{\Phi},
R\cos{\Theta}). 
\]
Then the conformal factor is $\Omega^{2}=R^2/r^2$, and the
embedding is given by
\begin{eqnarray}
\label{cembedding}
&&R=h(r), \quad \Theta=\theta, \quad \Phi=\phi,\\
&&T=\frac{4m}{r}h(r)\sqrt{1-2m/r}\sinh{(t/4m)}, 
\quad
X=\frac{4m}{r}h(r)\sqrt{1-2m/r}\cosh{(t/4m)}\quad \mbox{for}\quad r\geq 2m\nonumber\\
&&T=\frac{4m}{r}h(r)\sqrt{2m/r-1}\cosh{(t/4m)}, 
\quad
X=\frac{4m}{r}h(r)\sqrt{2m/r-1}\sinh{(t/4m)}\quad \mbox{for}\quad 0<r< 2m\nonumber
\end{eqnarray}
with
\be
\label{eq7}
h(r)=\exp\Big({\int \frac{p(r)}{q(r)} dr}\Big)
\ee
where 
\be
\label{pq}
p=48m^3-16m^2r-r^{3/2}\sqrt{r^3+2mr^2+4m^2r+72m^3}, \quad q=(32m^3-16m^2r-r^3)r.
\ee
The cubic $q$ has two imaginary roots, and one real root. The function $p$ has two  real roots: one negative   and one positive. 
The positive root of $p$ coincides with the real root of $q$, and
is given by
\be
\label{real_root}
\bar{r}=\frac{2}{3}\,\sqrt [3]{54+6\,\sqrt {129}}-{\frac {8}{\sqrt [3]{54+6\,\sqrt 
{129}}}}\sim 1.694 m.
\ee
Expanding $p$ and $q$ in $r$ around this root, and taking the limit we find
that the integrand in $h$ is regular at $\bar{r}$, and given by
\[
{\frac { \left( 237\,\sqrt {129}-1677 \right) \sqrt [3]{54+6\,\sqrt {
129}}+6192+ \left( -19\,\sqrt {129}+645 \right)  \left( 54+6\,\sqrt {
129} \right) ^{2/3}}{24768\,m}}\sim 0.88 m^{-1}.
\]
Therefore the conformal embedding of Schwarzschild in $\R^{4, 1}$ extends thorough the horizon all the way to the singularity $r=0$. 
The light cone of the origin in $\R^{4, 1}$  intersects the image of $M$ at a three--dimensional surface
$r=\bar{r}$, where $\bar{r}$ is given by (\ref{real_root}). Indeed
\[
0=T^2-X^2-R^2=\Omega^2\frac{32m^3-16m^2r-r^3}{r}.
\]
Moreover
\be
R\sim r, \quad \mbox{and}\quad \Omega^2\sim 1\quad\mbox{as}\quad r\rightarrow
\infty.
\ee
The plot of the conformal factor $\Omega=h(r)/r$ 
as a function of $r$ with $m=1$ is given below
 \begin{center}
  \label{fig1_paul}
  \includegraphics[width=5cm,height=5cm,angle=0]{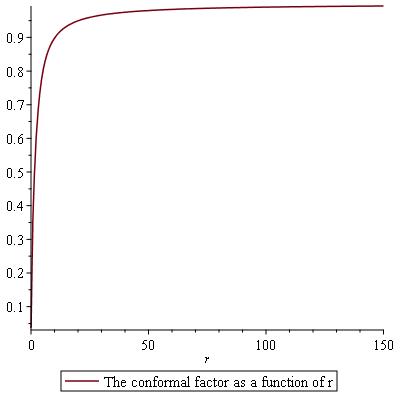}
\begin{center}
{{\bf Figure 1.} Conformal factor}
\end{center}
   \end{center}
In the case of the Reissner--Nordstr\"om metric with
\[
V=1-\frac{2m}{r}+\frac{Q^2}{r^2}
\]
a choice of $a$ can be made to extend the embedding (\ref{h_eq})
through the outer horizon $r_+=m+\sqrt{m^2-Q^2}$, but it then becomes singular at the inner horizon.
The embedding breaks down at the horizon of the extreme RN with $m=Q$ - we shall return extreme RN in
\S\ref{time_emb}, where we consider some time dependent embeddings.
\subsection{Embedding in Kruskal coordinates.}\label{ekc}
Set $s=r/m$, then
\begin{eqnarray*}
&&32s^{-3}e^{-s/2}dudv=R^{-2}(d(T-X)d(T+X)-dR^2),\\
&&\mbox{where}
\quad uv=K(s)=(1-s/2)e^{s/2}.
\end{eqnarray*}
Set $T-X=ue^{k(s)}, T+X=ve^{k(s)}, R=h(s)$. Then
\[
e^{2k}/h^2=32 s^{-3}e^{-s/2}, \quad
e^{2k}/h^2(K'k'+K(k')^2)-(h'/h)^2=0.
\]
We solve the first equation for $k$, and substitute to the 2nd equation which is now
an ODE for $h'/h$. The solution is
\begin{eqnarray*}
T+X&=& 4\sqrt{2}\Big(\frac{m}{r}\Big)^{3/2}e^{-r/4m} h(r) v\\
T-X&=& 4\sqrt{2}\Big(\frac{m}{r}\Big)^{3/2}e^{-r/4m} h(r) u
\end{eqnarray*}
where outside the horizon
\[
u=-\sqrt{r/2m-1}e^{(-t+r)/4m}, \quad v=
\sqrt{r/2m-1}e^{(t+r)/4m}.
\]
This, not surprisingly, agrees with (\ref{eq7}).
\subsection{The $r=0$ singularity}
Note that
\[
X^2-T^2=16m^2\frac{h(r)^2}{r^2}\Big(1-\frac{2m}{r}\Big)\quad\mbox{for}\quad 
r>0
\]
so that, in particular $X^2-T^2\rightarrow 16m^2$ as $r\rightarrow\infty$.
Consider the limit $r\rightarrow 0$ instead. Near $r=0$
(\ref{eq7}) gives
\[
\frac{p}{q}=\frac{3}{2r}+\frac{1}{4m}+O(\sqrt{r}), \quad\mbox{so}
\quad h(r)\rightarrow r^{3/2}
\]
and
\[
(T, X, R)\rightarrow \Big(4\sqrt{2}m^{3/2}\cosh{(t/4m)}, 
4\sqrt{2}m^{3/2}\sinh{(t/4m)}, 0\Big)\quad \mbox{as}\quad r\rightarrow 0.
\]
Thus the Schwarzschild singularity is mapped to the hyperbola
\[
T^2-X^2=32m^3, \quad X_2=X_3=X_4=0
\]
in the five--dimensional Minkowski space. The conformal factor 
$\Omega^{-2}=r^2/h(r)^2$ in (\ref{5-met}) blows up like $1/r$.
\section{Causality and Scri}
\label{seccs}
In this Section we shall study the conformal embedding of Proposition \ref{prop1} in the context of Penrose's conformal infinity, and find that 
the images of the future and past conformal infinities of the compactified
Schwarzschild space--time are points on the future and past conformal 
infinities of the ambient five--dimensional Minkowski space.

Let $p, q$ be points  in $M$ such that $q$ is in the causal future of $p$, which we denote by $p\prec q$. It follows that
$\iota(p)\prec\iota(q)$ with respect to the causal structure of the 5D Minkowski space. 
To see it consider a time-like curve  $\gamma\subset M$ containing $p, q$ with a tangent vector field
$V\subset \Gamma(TM)$. Thus $g(V, V)>0$. However
\[
g(V, V)=\Omega^{-2}\eta_{\mu\nu}V^aV^b\frac{\p X^{\mu}}{\p x^a}                   \frac{\p X^{\nu}}{\p x^b}
=\iota^{*}(\Omega^{-2}\eta(\iota_*V, \iota_*V))
\]
so the image of $\gamma$ is also time--like.

\begin{prop}
\label{prop2}
Let $(\II_\pm)^{Schw}$ and $(\II_\pm)^{5}$ be asymptotic null infinities of the compactified
Schwarzschild manifold $\overline{M}$, and the compatified $(4+1)$ dimensional Minkowski space
$\overline{\R}^{4, 1}$ respectively. The conformal embedding of Proposition \ref{prop1} extends to a map
$\iota:\overline{M}\rightarrow  \overline{\R}^{4, 1}$ such that $\iota( (\II_\pm)^{Schw})=p_\pm$ where
$p_-\in (\II_-)^{5}$ and $p_+\in (\II_+)^{5}$ are points with coordinates $(0, N)$, where $N\subset S^3$ is the north pole.
\end{prop}
{\bf Proof.}
Set
\[
\rho=\sqrt{X^2+R^2}, \quad R=\rho\sin{\psi}, \quad X=\rho\cos{\psi}
\]
so that
\[
-dT^2+dX^2+dR^2+R^2\gamma_{S^2}=-dT^2+d\rho^2+\rho^2(d\psi^2+\sin^2{\psi}
\gamma_{S^2}),
\]
where $\gamma_{S^2}=d\theta^2+\sin{(\theta)}^2d\phi^2$.
Note that for $X>0$
\[
\rho=X\sqrt{1+(R/X)^2}.
\]
If $v$ is finite, and $u\rightarrow-\infty, r\rightarrow \infty,
t\rightarrow -\infty$ (which is the past null infinity $(\II_-)^{Schw}$ of Schwarzschild)
then 
\[
\frac{R}{X}=\frac{r}{4m}\frac{1}{\sqrt{1-2m/r}}\frac{1}{\cosh{(t/4m)}}\sim
\frac{1}{2m}r e^{t/4m}
\]
but (as $v$ is finite) $e^{t/4m}\sim \sqrt{2mr}e^{-r/4m}$ so that
\[
\frac{R}{X}\sim \frac{r\sqrt{r}}{\sqrt{2m}}e^{-r/4m}\rightarrow 0 \quad\mbox{as}\quad r\rightarrow \infty.
\]
Therefore
\[
V\equiv T+\rho\sim T+X\rightarrow 0
\]
and $\iota(\II_-)^{Schw}\subset (\II_-)^{5}$. Similarly
$\iota(\II_+)^{Schw}\subset (\II_+)^{5}$ is given by $U\equiv T-\rho=0$.
In this limit we also get $\psi\rightarrow 0$. 
Here $\II_\pm \cong \R\times S^3$ are future and past null infinities
of the Minkowski space $\R^{4, 1}$.

Setting
\[
{\tau}=\arctan{V}+\arctan{U}, \quad \chi= \arctan{V}-\arctan{U}
\]
the Minkowski metric on $\R^{4, 1}$ is conformal to the Einstein cylinder
\[
\hat{\eta}=d{\tau}^2-d\chi^2-\sin^2{(\chi)}\gamma_{S^3}
\]
and the image of $(\II_-)^{Schw}$ is $\tau=-\pi/2, \chi=\pi/2$ which is an equatorial
$S^3$ in $S^4$. However $\psi=0$ is a north pole on $S^3$, so we get a 
point $p_1$ with coordinates $(V=0, \psi=0)$ on $({\II}_-)^5$.
Similarly $({\II_+})_{Schw}$ maps to $p_+$ with coordinates $(U=0, \psi=0)$ 
on $({\II}_+)^5$.
  \begin{center}
  \label{rog_1}
  \includegraphics[width=6cm,height=6cm,angle=0]{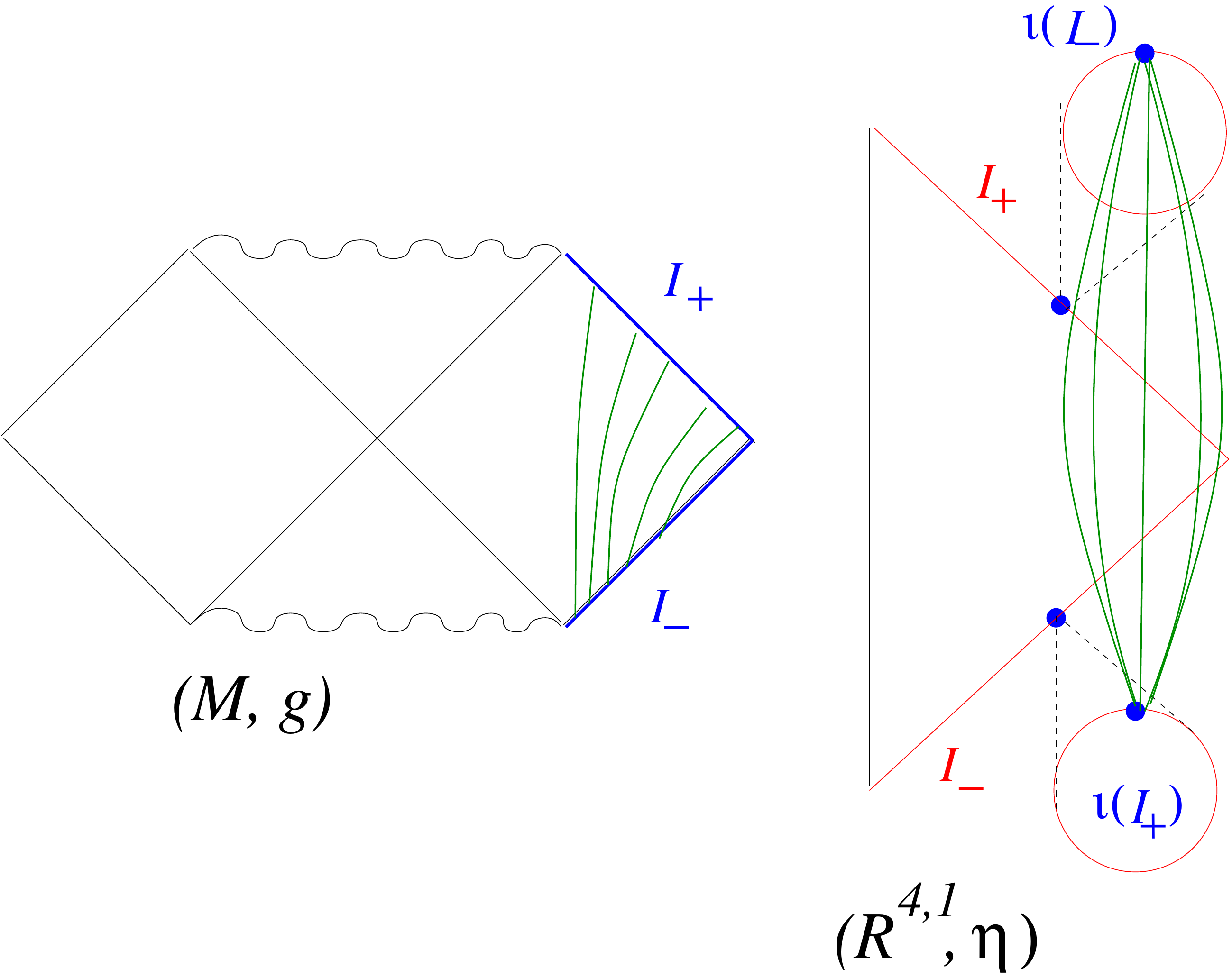}
\begin{center}
{{\bf Figure 2.} Conformal lift of Scri.}
\end{center}
  \end{center}
  \koniec
\section{Penrose's Schwarzchild causality}
\label{seccal}
If it {\em were} possible to construct a quantum theory of gravity as a 
Poincar\'e--invariant expansion around the 
flat Minkowski metric $\eta$, then the causal relations of the perturbed metric 
$g=\eta+\epsilon \eta_1+\epsilon^2\eta_2+\dots$  should agree with causal relations of  $\eta$ in that the time--like
curves with respect to $g$ should also be time--like with respect to $\eta$. Lets write this condition as
$g<\eta$. If this condition does not hold, then there would exist fields propagating inside the $g$ light--cones
which are  tachyonic with respect to the $\eta$ light--cones. According to  the standard rules of QFT these
fields would correspond to non--commuting operators on $(M, g)$. But this would imply that these operators
are also non--commuting for some space--like separated points on $(\R^{3,1}, \eta)$ which is impossible.
Penrose's argument \cite{roger_paper} shows that for the Schwarzschild metric the condition $g<\eta$  fails asymptotically.
We shall review the argument below, and argue that our findings about the 
conformal embedding agree with results of \cite{roger_paper}.
\vskip 5pt

In \cite{roger_paper} Penrose considers two (equivalent) properties
of some asymptotically flat space--times, and shows that they hold for
the compactified Schwarzschild space--time, but not for the compactified 
Minkowski space--time.
\begin{itemize}
\item[{\bf P1}] Let $(a_-, b_+)\in \II_-\times\II_+$ be any pair of points. Then $a_-\ll b_+$, i. e. $\exists$ a future directed
time--like curve from $a_-$ to $b_+$.
\item[{\bf P2}] If $\alpha$ and $\beta$ are endless time-like curves in $(M, g)$ then $\exists (a\in \alpha, b\in \beta)$
such that $a\ll b$.
\end{itemize}
First let us see that {\bf P1} fails for the Minkowski metric $\eta=d{X_0}^2-d{X_1}^2-\dots - d{X_D}^2$. Consider
two branches $h_1$ and $h_2$ of the  hyperbola ${X_1}^2-{X_0}^2=1, X_2=X_3=\dots X_D=0$. Both $h_1$ and $h_2$ are time--like in $\R^{D, 1}$ and yet  any pair of points $a\in h_1, b\in h_2$ are space separated. Let $a_-$ be the end--point
of $h_1$ on $\II_-$ and $b_+$ be the end--point of $h_2$ on $\II_+$. Then $b_+$ does not belong to the chronological future of $a_-$, and computing the angles on asymptotic spheres 
$X_0+\sqrt{{X_1}^2+\dots +{X_D}^2}=const$ in $\II_-$ and $X_0-\sqrt{{X_1}^2+\dots +{X_D}^2} =const$ in $\II_+$ we find that $a_-$ and $b_+$ are antipodal points on these spheres. 
\vskip 5pt

Penrose then argues that {\bf P1} holds for the Schwarzschild space--time in $3+1$ 
dimensions\footnote{This property also holds in $2+1$--dimensions, where the metric is locally flat but admits a conical singularity, 
but it does not appear to hold in higher dimensions \cite{CD19}.}.
Consider a geodesic Lagrangian for the Schwarzschild metric
\[
{\mathcal L}=\frac{1}{2}\Big(V\dot{t}^2-V^{-1}\dot{r}^2-r^2(\dot{\theta}^2+\sin^2\theta\dot{\phi}^2)\Big), \quad
V=1-\frac{2m}{r}
\]
where $\cdot=d/d\tau$ and $\tau$ is an affine parameter. Let $\gamma$ be a null geodesic such that
\be
\label{boundary_cond}
r=r_0,\quad \theta=\phi=\pi/2, \quad  \dot{r}=\dot{\theta}=0 \quad
\mbox{at}\quad t=0.
\ee
We can normalise $\tau$ so that $\tau=0$ at $t=0$, and
\[
\frac{\p {\mathcal L}}{\p \dot{t}}=V\dot{t}=1.
\]
The null condition gives
\be
\label{null}
V^{-1}(\dot{r}^2-1)+\frac{A^2}{r^2}=0, \quad\mbox{where}\quad A=\frac{\p \mathcal L}{\p \dot{\phi}}=r^2\dot{\phi}=\mbox{const}.
\ee
Evaluating this at $t=0$ for $\dot{\phi}>0$ yields
\[
A=\frac{r_0}{\sqrt{V(r_0)}}.
\]
Solving the condition (\ref{null})  for $t$ yields
\be
\label{t_formula}
t=\int_{r_0}^r \frac{V^{-1}(\rho)}{\sqrt{1-\frac{A^2}{\rho^2}V(\rho)}}d\rho.
\ee
Using the identity
\[
\int_{r_0}^r\Big(1-\frac{2m}{\rho}\Big)^{-1}d\rho=r+2m\ln{(r-2m)}-r_0-2m\ln{(r_0-2m)}
\]
now shows that the retarded time $\hat{u}=t-r-2m\ln{(r-2m)}$ can be
written as
\be
\label{roger}
\hat{u}=-r_0-2m r_0\ln{(r_0-2m)}+\int_{r_0}^r \chi(\rho) d\rho
\ee
where
\be
\label{chi}
\chi(\rho)=V(\rho)^{-1}\Big( \sqrt{1
-\frac{{r_0}^2}{\rho^2}\frac{V(\rho)}{V(r_0)}}^{-1}-1 \Big).
\ee
The null geodesic $\gamma$ will reach a point on $\II_+$ therefore the integral in (\ref{roger})
converges as $r\rightarrow\infty$ (which can also be verified directly). In \cite{roger_paper} Penrose applies some estimates
to the integral (which are valid if $r_0>5m$) and shows 
that\footnote{The proof  goes as follows:
\be
\label{roger_estim}
\int_{r_0}^{\infty}\chi(\rho)d\rho <\int_{r_0}^{A} \chi(\rho)d\rho+\int_A^{\infty} \psi(\rho)d\rho,
\ee
where
\[
\psi(\rho)= V(r_0)^{-1}
\Big(\sqrt{1-\frac{{r_0}^2}{\rho^2 V(r_0)}}^{-1}-1 \Big)>\chi(\rho) \quad\mbox{for}\quad  \rho>A\equiv
\frac{r_0}{\sqrt{V(r_0)}}.
\]
The second integral in (\ref{roger_estim}) can be evaluated explicitly to give $r_0V(r_0)^{-3/2}$ which tends
to $r_0$ for large $r_0$. The first integral can be bounded from above by a constant which does not depend
on $r_0$.}
\[
\int_{r_0}^{\infty} \chi(\rho)d\rho< r_0+\mbox{const}
\]
for large $r_0$. Therefore, from (\ref{roger})
\be
\label{limit_u}
\lim_{r_0\rightarrow\infty} \hat{u}=-\infty
\ee
which holds as long as $m>0$.
\vskip5pt

Chose an arbitrary pair of values $(\hat{u}_c, \hat{v}_c)$. The argument  leading to (\ref{limit_u})
shows that $\exists r_0(\hat{u}_c)$ s.t. a null geodesic satisfying the initial condition (\ref{boundary_cond})
reaches $\II_+$ at some $\hat{u}_+<\hat{u}_c$. This null geodesic will, before reaching $\II_+$, meet an outgoing radial null geodesic $\beta$ given by
\[
\hat{u}=\hat{u}_0, \quad \theta=\pi/2, \quad \phi=\phi_0=\mbox{const}
\]
where $\phi_0$ is any angle in the range $[\pi/2, \pi]$. As $\hat{u}$ is increasing along $\gamma$ we must have $\hat{u}_0<\hat{u}_+$.

Applying the same argument to $\II_-$ shows that for any choice of $\hat{v}_c$ there exists $r_0(\hat{v}_c)$ such that $\gamma$ reaches
$\II_-$ at some $\hat{v}_+>\hat{v}_c$. Pick $r_0=\mbox{max}(r_0(\hat{u}_c), 
r_0(\hat{v}_c))$. Again, $\gamma$ will meet an incoming  radial null
geodesic $\alpha$
\[
\hat{v}=\hat{v}_1, \quad \theta=\pi/2, \quad \phi=\phi_1=\mbox{const},
\]
where now $\phi_1$ is any value in $[0, \pi/2]$
and $\hat{v}=t+r+2m\ln{(r-2m)}$. Let 
$\widehat{\alpha\gamma\beta}$ be a null geodesic 
which consist of three segments:
from $(\hat{v}_1, \pi/2, \phi_1)$ on $\II_-$ along $\alpha$, then from the meeting point of $\alpha$ and $\gamma$ along $\gamma$ and finally from 
the meeting point of $\gamma$ and $\beta$ along $\beta$ and up to $(\hat{u}_0, \phi_0, \theta=\pi/2)$ on $\II_+$. This null geodesic
can be smoothen to give a time-like curve, and by a rotation of the $(\theta, \phi)$ coordinates on $S^2$ any point on $\II_-$ can be connected to any point on $\II_+$.
  \begin{center}
  \label{rog_3}
	  \includegraphics[width=8cm,height=5cm,angle=0]{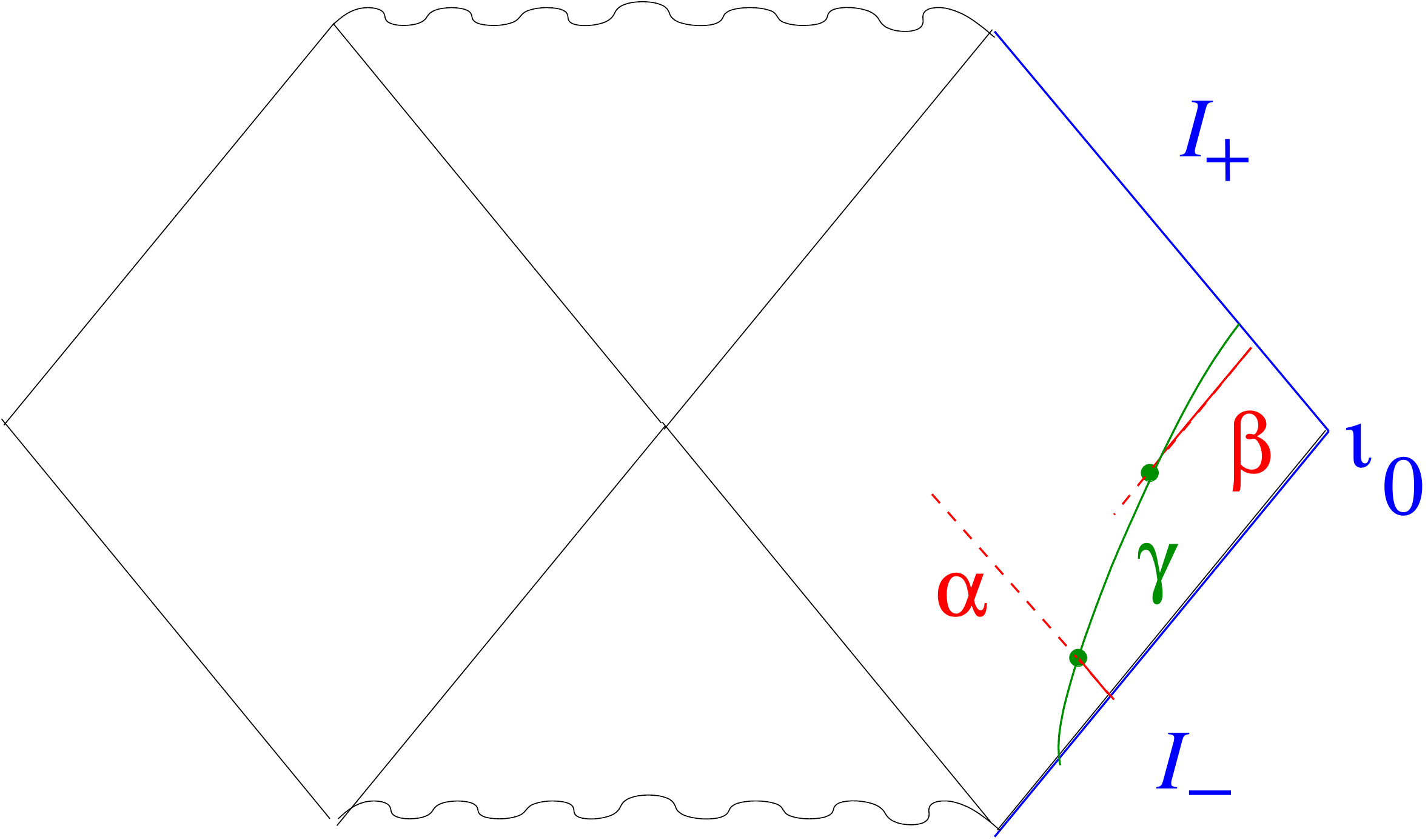},
\begin{center}
{{\bf Figure 3.} Schwarzschild causality.}
\end{center}
 \end{center}
In particular antipodal points can also (and unlike in the Minkowski space) be connected.
\vskip5pt

Our findings (Prop \ref{prop2}) about the image of $\II_\pm$ of the Schwarzschild space--time under the conformal embedding 
(\ref{cembedding}) agree\footnote{We are unable to deduce  Penrose's  result using our embedding, as although
$a\ll b$ implies $\iota(a)\ll \iota(b)$  it can still be the case that 
$b\notin I_+(a)$  but $\iota(a)\ll\iota(b)$ if the time--like curve joining  $\iota(a)$ and $\iota(b)$ is not an image of a curve in $M$.} with Penrose's result.  By \cite{roger_paper}  any $(a_-, b_+)\in\II_-\times\II_+$ are chronologically  related i.e.
$a_-\ll b_+$, therefore it should be the case that $\iota(a_-)\ll\iota(b_+)$ in $\R^{4,1}$. We have found that
$\iota(a_-)=(V=0, N)$ and $\iota(b_+)=(U=0, N)$, where $N\in S^3$ is the north pole corresponding to $\psi=0$.
These two points in $\bar{\R^{4, 1}}$ are end points of a time--like curve which is {\em one} branch of the hyperbola
${X_1}^2-{X_0}^2=16m^2, X_2=X_3=X_4=0$, so are indeed causally related.

\section{Hawking to Unruh}
\label{htou}

We have so far focused on the geometric aspects of conformal embeddings. The emphasis has been on Lorentzian (rather than Riemannian)
examples which has prepared the ground for exploring applications in physics.  The physical effects (classical or quantum) induced
by conformal curvature of a  Lorentzian manifold should have their counterparts in the flat ambient Lorentzian space. We expect this
correspondence to extend only to conformally invariant effects, and in this section we shall argue that the Hawking effect gives one example.

The Hawking radiation \cite{Hawking} is a kinematical effect. It does not depend on the
Einstein equations, but only on an existence of a Lorenzian metric with a horizon. The Hawking temperature measured by asymptotic observers is given by ${\T}_H=\kappa/2\pi$, where
$\kappa$ is the surface gravity of a Killing horizon of some Killing 
vector $K$ defined 
by

\[
\nabla_a(|K|^2)=-2\kappa K_a.
\]
This surface gravity is invariant under conformal rescallings
$g\rightarrow \hat{g}=\Omega^{2}g$ as long as the conformal factor and its gradient
are regular on the horizon (see \cite{jacobson} where other, equivalent
definitions of surface gravity are discussed in the context of conformal rescallings), and $\Omega\rightarrow 1$ as $r\rightarrow\infty$ as then
the normalisation $g(\p_t, \p_t)=\hat{g}(\p_t, \p_t)\rightarrow 1$ is preserved.
\begin{eqnarray*}
-2\hat{\kappa}\hat{g}_{ab}K^b&=& \hat{\nabla}_a(\hat{g}_{bc}K^bK^c) =\Omega^{2}\nabla_{a}(g_{bc}K^bK^c)+2{|K|_g}^2\Omega\nabla_{a}\Omega\\
&=&-2\Omega^{2}\kappa K_a
\end{eqnarray*}
as $g_{ab}K^aK^b=0$ on the horizon. Therefore $\kappa=\hat{\kappa}$.
It has been argued (see e.g. \cite{jacobson} and
\cite{nielsen}) that the original derivation of Hawking based on the Bogoliubov coefficients should also lead to a conformally invariant temperature.

We shall show that under the isometric embedding (\ref{cembedding}) of
the conformally rescalled Schwarzschild metric in flat $\R^{4,1}$, the Hawking temperature maps to the Unruh
temperature measured by the accelerating observer. Our procedure
is analogous to that of Deser and Levin \cite{deser} (see also \cite{pestun1}), who  mapped the
Hawking temperature of the Schwarzschild metric to the Unruh temperature
of its Fronsdal embedding
in the flat six--dimensional Lorenzian space \cite{fronsdal}.
Consider a curve in the flat Minkowski space $\R^{4, 1}$ parametrised by $t$, and given by (\ref{cembedding}) with $\theta, \phi, r$ fixed.  This
gives $X_3, X_4, X_5$ constants, and
\[
{X}^2-T^2=\frac{16m^2\;h(r)^2}{r^2}\Big(1-\frac{2m}{r}\Big)\equiv \alpha^{-2}, 
\quad\mbox{where $r$ is fixed.}
\]
This is a worldline of an accelerating observer  moving along a hyperbola in the flat 5--dimensional Minkowski space
with constant acceleration $\alpha$.

This observer is experiencing the Unruh temperature \[\T=\alpha/2\pi.\] 
The observer follows a trajectory of a boost\footnote{By a theorem of
\cite{tod_circles} a trajectory of a (non--null) hyper--surface orthogonal Killing vector is a conformal geodesic. The magnitude $a_M$ of the acceleration is constant if $(M, g)$ is Einstein. Consider this Killing vector to be $V=\p/\p t$. Its integral curve
$\gamma\subset M$ lifts to a constant acceleration hyperbola in the Minkowski space
$(\R^{4, 1}, \eta)$ (the embedding is non--isometric), or to a curve in
$(\hat{\R}^{4, 1}, \Omega^{-2}\eta)$ with acceleration $a_5$. 

In general, if the particle trajectory be tangent to an affinely parametrised timelike vector $V\in TM$. The acceleration in the embedding space is
\be
\label{accel}
\frac{d U}{d\lambda}=\nabla_U U+K(U, U)
\ee
where $K$ is the second fundamental form of the embedding 
Squaring (\ref{accel}) gives
\[
(a_5)^2=(a_M)^2+|K(U, U)|^2.
\]
For the isommetric embedding in conformally flat $\R^{4, 1}$ the intrinsic acceleration
$a_M$ is constant, but $a_5$ is not constant, and the contribution comes from $K$.
If we instead isometrically embed  $\Omega^{2}g$ in the flat $\R^{4, 1}$, then
$a_M$ is not constant. 
For the flat Minkowski space $\eta_{\alpha\beta}U^{\alpha}U^{\beta}=-1$ and  $A^{\alpha}=dU^{\alpha}/d\lambda$. 
Consider a curve
$T(\lambda), X=X^1(\lambda), X^2=const, X^3=const, X^4=const$. Then  
\[
dT/d\tau=U^0, dX/d\tau =U^1, \eta(U, A)=0, \eta(A, A)=a^2=const
\]
and we find $X=a^{-1} \cosh(a\tau), T=a^{-1} \sinh(a\tau)$.
In general, the coordinate transformation
$X=a^{-1}x \cosh(a\tau), T=a^{-1}x \sinh(a\tau)$
gives $dT^2-dX^2=-a^{2}dx^2+x^2dt^2$ and the curves $x=1$ have constant acceleration
$a$.}
in $\R^{4, 1}$ which in our case (at least in the image
of a region in $\hat{M}$ outside the horizon $r>2m$) is a push forward
$K$ of $\p/\p t$ by the embedding map. An observer at any other value of $r$ (say $r=r_0$) will experience
a temperature $\T_0$ which is related to $\T$ by Tolman's law
\[
|K|_{\eta}(r) \T=|K|_{\eta}(r_0)\T_0, \quad\mbox{where}\quad {|K|_\eta}^2=\eta(K, K).
\]
In our case the flat Minkowski metric $\eta=\eta^{\mu\nu}dX_\mu dX_\nu$ restricted to the curve (\ref{cembedding}) is 
\[
\eta=\frac{h(r)^2}{r^2}\Big(1-\frac{2m}{r}\Big)dt^2.
\]
Therefore $|K|_{\eta}(r)=(h/r)\sqrt{1-2m/r}$.
Taking $r_0\rightarrow\infty$ we get $|K|_{\eta}(r_0)\rightarrow 1$
so that
\[
\T_0=\frac{1}{8\pi m}
\]
and the Unruh temperature measured by observers at infinity
in $\R^{4, 1}$ agrees with the Hawking temperature\footnote{
Note that this result does not apply to observers in the Schwarzschild space time who do not follow a trajectory of a time--like Killing vector.
There are other possibilities, e.g. free falling observers,  where the temperature measured by an observer differs from the Hawking 
temperature \cite{island}.} 
$\hat{\T}_H=\hat{\kappa}/2\pi$ of $(M, \hat{g})$.
However, as the conformal factor $\Omega^{2}$ and its gradient 
are both regular at the Killing horizon of $\p/\p t$, this is also the Hawking temperature of the Schwarzschild black hole. 

The conformal invariance of the Hawking temperature is also in agreement with
Euclidean quantum gravity, where the Hawking temperature is the quarter of the 
period of the imaginary time direction\cite{euclidean_qg}, where the periodicity makes
the Schwarzschild metric regular at $r=2m$, and the domain of $r$ is restricted
to $r>2m$. This period is unchanged if the metric is rescaled by the conformal factor $\Omega^2$, as long as $\Omega$ is regular at $r=2m$. The formula
(\ref{eq7}) with $\Omega=h(r)/r$ gives
\begin{eqnarray*}
\ln{\Omega}&=&\int_{r}^{\infty}\Big(\frac{1}{\rho}-\frac{p(\rho)}{q(\rho)}\Big)d\rho\\
&=&\int_{r/2m}2\frac{\sqrt{1+x^{-1}+x^{-2}+9x^{-3}}-1-2x^{-3}}{x(4x^{-3}-4x^{-2}-1)}dx.
\end{eqnarray*}
Computing the last integral  numerically from $1$ to $\infty$ gives
\[
\Omega(2m)\sim 0.576
\]
which does not depend on the mass $m$.
\section{Time dependent embeddings}
\label{time_emb}
The conformal embedding of Proposition \ref{prop1} is time independent in the sense that the conformal factor
$\Omega$ is constant along the time--like static Killing vector of $(M, g)$. In this section we shall construct
some time--dependent embeddings. In the proof of Proposition \ref{prop1} we demonstrated that a 
spherically symmetric (but possibly time dependent) conformal embedding of (\ref{eeq1})  
into $\R^{4, 1}$ arises from an isometric embedding
of 
\[
g_2=
\frac{1}{r^2}(V^{-1}dr^2-Vdt^2)
\]
into a patch of $AdS_3$ with the metric
\[
G_3=\frac{dR^2+dX^2-dT^2}{R^2}.
\]
We can  make use of time--dependent isometries of $AdS_3$ co construct
time dependent embeddings. For example a one--parameter family of isometries
\be
\label{dsboost}
R_c=\frac{R}{Bc^2+2Xc+1}, \quad 
T_c=\frac{T}{Bc^2+2Xc+1}, \quad X_c=\frac{X+Bc}{Bc^2+2Xc+1},
\ee
of $G_3$  (where $B=R^2+X^2-T^2$) is generated by the Killing vector field
\[
K=2X(R\p_R+T\p_T)-(R^2-X^2-T^2)\p_X.
\]
Taking  as in $(X, R, T)$ are given by (\ref{eeq1}) and (\ref{h_eq}) we find that
\[
G_c=\Omega^{-2}(dT_c^2-dX_c^2-dR_c^2-R_c^2 (d\theta^2+\sin{\theta}^2d\phi^2)).
\]
with $\Omega^2=R_c^2/r^2$ pulls back to (\ref{four_m}), but now the conformal embedding is time--dependent.
\subsection{Extreme  Reissner--Nordstr\"om metric}
To construct a different time dependent embedding
go back to (\ref{5-met}) and set
\[
\Omega^{2}=V^{-1}, \quad
T=t, \quad R=r\sqrt{V}^{-1}, \quad
X=\int\frac{\sqrt{4-V^{-1}(2V-rV')^2}}{2V}dr.
\]
Then $G$ pulls back to (\ref{four_m}).
For the Schwarzschild metric we find
\[ 
X=\int\sqrt{\frac{rm(4r-9m)}{r-2m}}\frac{1}{r-2m}dr,
\]
which does not go through the horizon\footnote{The term $(4r-9m)$ is reminiscent of the Misner--Sharp
mass of the Schwarzschild metric conformally rescaled to the ultra-static form.}.
For the extreme RN we end up with elementary functions: 
\[
V=\Big(1-\frac{Q}{r}\Big)^{2}, \quad 
X=\frac{4\sqrt{Q}(r-2Q)}{\sqrt{r-Q}}.
\]
We can combine this embedding with a conformal inversion
\be
\label{inversion}
(\hat{X}, \hat{R}, \hat{T})=\Big(\frac{X}{X^2+R^2-T^2}, 
\frac{R}{X^2+R^2-T^2}, \frac{T}{X^2+R^2-T^2}\Big),
\ee
which is the combination of (\ref{dsboost}) with two  translations in the $X$--direction.
The resulting metric $\hat{R}^{-2}(d\hat{X}^2+d\hat{R}^2-d\hat{T}^2)$ is still isometric
to $r^{-2}(-Vdt^2+V^{-1}dr^2)$, but now the coordinates are regular (and in fact vanish)
at the extreme horizon. 
There are however other singularities which depend on $t$. 

In the near--horizon limit the extreme $RN$ metric reduces to the 
Bertotti--Robinson solution $AdS_2\times S^2$. To take this limit set
\[
r=Q\Big(1+\frac{\epsilon}{y}\Big)
\]
and apply the inversion (\ref{inversion}) to the rescalled coordinates
$(\epsilon T, \epsilon R, \epsilon X)$. Up to the linear terms in $\epsilon$ this gives
\[
\hat{T}=\frac{t}{Q(y^2-t^2)}-\epsilon \frac{20ty}{Q(t^2-y^2)^2} , \quad\hat{R}=\frac{y}{Q(y^2-t^2)}-\epsilon \frac{2t^2+18 y^2}{Q(t^2-y^2)^2}, 
\quad \hat{X}=\sqrt{\epsilon}\frac{4\sqrt{y}}{Q(y^2-t^2)}
\]
and
\[
G=\Omega^{-2}(d\hat{T} -d\hat{X}^2-d\hat{R}^2-\hat{R}^2 (d\theta^2+\sin{\theta}^2d\phi^2)), \quad\Omega^2=\hat{R}^2/r^2
\]
pulls back to
\[
g=Q^2\Big(\frac{dt^2-dy^2}{y^2} -d\theta^2-\sin{\theta}^2d\phi^2 \Big)
-\epsilon\frac{2Q^2}{y}\Big(\frac{dt^2+dy^2}{y^2} +d\theta^2+\sin{\theta}^2d\phi^2 \Big).
\]
In the limit $\epsilon=0$ this yields $AdS_2\times S^2$, where the horizon has been mapped to $y=\infty$
of $AdS_2$. Curiously the term first order in $\epsilon$ is proportional to a Riemannian product metric
on $\HHH^2\times S^2$.
\begin{appendix}
\section{Proof of Proposition \ref{prop_real}}
We know from Proposition \ref{propcomplex} that in Lorentzian
signature the reality of the invariants $(I, J)$ given by (\ref{IJ})
is a  necessary conditions for the existence of solutions to (\ref{em4}).
We shall now show that these conditions are also sufficient
for the existence of solutions to  (\ref{em4}) which give rise to real
second fundamental forms.

Choose a normalised spinor dyad $(o^A,\iota^A)$ and expand
\bea\label{1}
\sigma_{ABA'B'}&=&Xo_Ao_Bo_{A'}o_{B'}-2\overline{U}o_Ao_Bo_{(A'}\iota_{B')}+\overline{V}o_Ao_B\iota_{A'}\iota_{B'}\\
&&-2Uo_{(B}\iota_{B)}o_{A'}o_{B'}+4Yo_{(A}\iota_{B)}o_{(A'}\iota_{B')}-2\overline{W}o_{(A}\iota_{B)}\iota_{A'}\iota_{B'}\nonumber\\
&&+V\iota_A\iota_B o_{A'} o_{B'}-2W\iota_A\iota_Bo_{(A'}\iota_{B')}+Z\iota_A\iota_B\iota_{A'}\iota_{B'}.\nonumber
\eea
Note $X,Y,Z$ are real, $U,V,W$ are complex. Substitute into (\ref{em4}) and take components

\begin{subequations}
\begin{align}
-2\epsilon\psi_4&=2X\overline{V}-2\overline{U}^2\label{ap3}\\
8\epsilon\psi_3&=8Y\overline{U}-4X\overline{W}-4U\overline{V}\label{ap4}\\
-12\epsilon\psi_2&=2XZ-8Y^2+2V\overline{V}+8U\overline{W}-4\overline{U}W
\label{ap5}\\
8\epsilon\psi_1&=8YW-4ZU-4V\overline{W}\label{ap6}\\
-2\epsilon\psi_0&=2VZ-2W^2\label{ap7}.
\end{align}
\end{subequations}

We will solve these equations by cases. 
It is  worth noting the expression
\be\label{i1}
I=2\psi_0\psi_4-8\psi_1\psi_3+6\psi_2^2,\quad
J=6\psi_0\psi_2\psi_4+ 12 \psi_1\psi_2\psi_3-6{\psi_2}^3
-6{\psi_1}^2\psi_4-6\psi_0{\psi_3}^2
\ee
\subsubsection*{Type N}
We can choose the spinor dyad so that 
$\psi_1=\psi_2=\psi_3=\psi_4=0,\;\;\psi_0\neq 0$, 
and in particular $I=J=0$ so both are real. 
Solutions exist with $U=X=W=0$, $Y,V,Z$ nonzero and
$V\overline{V}=4Y^2,\;\;\;VZ=-\epsilon\psi_0.$
\subsubsection*{Type III}
We can choose the dyad so that  
$\psi_0=\psi_2=\psi_3=\psi_4=0,\;\;\psi_1\neq 0,$
and again $I=J=0$. Solutions exist with $U=X=0$, others nonzero and
\[V=W^2/Z,\;\;Y=-|W|^2/(2Z),\;\;\epsilon\psi_1=-W^2\overline{W}/Z.\]
\subsubsection*{Type D}
We can choose the dyad so that 
$\psi_0=\psi_1=\psi_3=\psi_4=0,\;\;\psi_2\neq 0.$
Now 
\be
\label{ijap}
I=6\psi_2^2,\;\;J=-6\psi_2^3 \quad \mbox{  so that  }\psi_2=-J/I,
\ee
and reality of $I,J$ forces reality of $\psi_2$ (which eliminates 
the Kerr solution in agreement with Corollary \ref{kercol}).
There are solutions with $U=V=W=0$, $(X, Y, Z)$ all nonzero and
$-6\epsilon\psi_2=XZ-4Y^2,$
provided $\psi_2$ is real.
\subsubsection*{Type II}
We can choose the dyad so that 
$\psi_0=\psi_1=\psi_4=0$ and 
$
\psi_2\psi_3\neq 0.
$
We again find (\ref{ijap}) with real $\psi_2$
There are solutions with $X=U=0$, rest nonzero and chosen as follows: choose real $|W|^2/Z$, solve
\[4Y^2=6\epsilon\psi_2+|W|^4/Z^4\]
for real $Y$, which can be done for real $\psi_2$ and suitable $\epsilon$ or large enough $|W|^2/Z$; then solve
$2\epsilon\psi_1=W(2Y-|W|^2/Z$
for $W$ (when $Z$ follows).
\subsubsection*{Type I}
This is the general case but we can choose the dyad so that
\[\psi_0=0=\psi_4,\]
and assume $\psi_1\psi_2\psi_3\neq 0$ as other cases have been done already. These force
\[XV-U^2=0=ZV-W^2.\]
We can eliminate the possibility $V=0$ at once as this leads to $U=0=W$ and then $\psi_1=0=\psi_3$, a contradiction. With $V\neq 0$, if $X=0$ then also $U=0$ and $\psi_3=0$, also a contradiction so w.l.o.g. we have $XZUW\neq 0$. Now
\[X/Z=U^2/W^2\in\mathbb{R},\]
and we have a dichotomy: $U/W$ is real or imaginary. We consider the cases separately.
\subsubsection*{Type I, Case (a)}
If $U/W=r$ is real we have $U=rW$ and $X=r^2Z$. From (\ref{ap4})
\[2\epsilon\psi_3=2rY\overline{W}-rW\overline{V}-r^2Z\overline{W},\]
from (\ref{ap6})
\[2\epsilon\psi_1=2YW-\overline{W}V-rZW,\]
and from (\ref{ap5})
\be\label{s0}-6\epsilon\psi_2=r^2Z^2+V\overline{V}-4Y^2+2rW\overline{W}.\ee
Note that $\psi_2$ is real and $\psi_1=r\overline{\psi_3}$. These conditions imply but are stronger than reality of $I$ and $J$. 
From the vanishing of $\psi_4$ we have $V=W^2/Z$ so that
\be\label{s1}2\epsilon\psi_1=W(2Y-rZ-|W|^2/Z).\ee
We can solve the system as follows: introduce $a=\epsilon\psi_1/W$ which is real by (\ref{s1}) and solve (\ref{s1}) for $Y$
\be\label{s2}
Y=a+\frac12rZ+\frac12|W|^2/Z;
\ee
substitute into (\ref{s0}) to obtain a quadratic for $Z$:
\be\label{s3}
2arZ^2+(2a^2-3\epsilon\psi_2)Z+2a|W|^2=0;
\ee
this is real, as it must be, and there will be real solutions if the discriminant is positive; this is the condition
\[4a^4-12a^2\epsilon\psi_2+(9\psi_2^2-16\psi_1\psi_3)>0,\]
which certainly holds for large enough $a$.

Note that we do not need to worry about the value of $r$: the dyad 
$(o^A,\iota^A)$ has the scaling freedom
$(o^A,\iota^A)\rightarrow(\hat{o}^A,\hat{\iota}^A)=(\lambda o^A,\lambda^{-1}\iota^A)$
for arbitrary nonzero complex $\lambda$ and under this
$r\rightarrow\hat{r}=
(\lambda\overline{\lambda})^{-2}r,$
and we can always arrange $r=\pm 1$. 
\subsubsection*{Type I, Case(b)}
Now $U=irW$ and $X=-r^2Z$ for some real $r$. Also $V=W^2/Z$. From (\ref{ap4})
\be\label{b1}2\epsilon\psi_3=-ir\overline{W}(2Y+|W|^2/Z+irZ),\ee
from (\ref{ap6})
\be\label{b2}2\epsilon\psi_1=W(2Y-|W|^2/Z-irZ),\ee
and from (\ref{ap5})
\be\label{b3}-6\epsilon\psi_2=-r^2Z^2+|W|^4/Z^2-4Y^2+6ir|W|^2.\ee
From (\ref{b1}) and (\ref{b2}) we obtain two expressions for $Y$:
\be\label{b4}
Y=i\epsilon\frac{\psi_3}{r\overline{W}}-\frac{|W|^2}{2Z}-\frac{ir}{2}Z,
\ee
and
\be\label{b5}
Y=\epsilon\frac{\psi_1}{W}+\frac{|W|^2}{2Z}+\frac{ir}{2}Z.
\ee
Since $Y$ must be real, (\ref{b4}) gives
\be\label{b6}
irZ=\frac{i\epsilon\psi_3}{r\overline{W}}+\frac{i\epsilon\overline{\psi}_3}{rW},\ee
while (\ref{b5}) gives
\be\label{b7}
irZ=\frac{\epsilon\overline{\psi}_1}{\overline{W}}-\frac{\epsilon\psi_1}{W}.\ee
These agree only if
\[\frac{i\epsilon\psi_3}{r\overline{W}}+\frac{i\epsilon\overline{\psi}_3}{rW}=\frac{\epsilon\overline{\psi}_1}{\overline{W}}-\frac{\epsilon\psi_1}{W},\]
whence
\[\frac{W}{\overline{W}}=\frac{r\psi_1+i\overline{\psi}_3}{r\overline{\psi}_1-i\psi_3}.\]
From (\ref{b3}) we find
\[-\epsilon(\psi_2-\overline{\psi}_2)=2ir|W|^2,\]
(in type I we are assuming $XZUVW\neq 0$ so in this case we cannot have $\psi_2$ real -- that has to be case (a)) so that
\be\label{b8}r|W|^2=\frac{i\epsilon}{2}(\psi_2-\overline{\psi}_2),\ee
and $W=|W|e^{i\omega}$ with
\be\label{b9}e^{2i\omega}=\frac{r\psi_1+i\overline{\psi}_3}{r\overline{\psi}_1-i\psi_3}.\ee
We obtain $Z$ from either (\ref{b6}) or (\ref{b7}) (these are now equivalent) as
\be\label{b10}
Z=\frac{\epsilon}{r|W|}\frac{(\psi_1\psi_3+\overline{\psi}_1\overline{\psi}_3)}{|r\overline{\psi}_1-i\psi_3|}.\ee
For $Y$, (\ref{b4}) and (\ref{b5}) are now both real. We can add them to obtain
\be\label{b11}
Y=\frac{\epsilon}{2r|W|}\frac{(r^2\psi_1\overline{\psi}_1-\psi_3\overline{\psi}_3)}{|r\overline{\psi}_1-i\psi_3|},\ee
but the difference will give a constraint. This turns out to be
\be\label{b11a}(\psi_2-\overline{\psi}_2)^2=4(\psi_1\psi_3+\overline{\psi}_1\overline{\psi}_3)\ee
which rearranges as
\be\label{b12}
I+\overline{I}=4(\psi_2^2+\overline{\psi}_2^2+\psi_2\overline{\psi}_2).
\ee
This is therefore a necessary condition on Case (b). There is a quicker route to it: multiply (\ref{b1}) and (\ref{b2}) and make use of (\ref{b3}) to obtain
\[4\psi_1\psi_3=(\psi_2-\overline{\psi}_2)(2\psi_2+\overline{\psi}_2),\]
which rearranges as
\be\label{b13}I=2(\psi_2^2+\overline{\psi}_2^2+\psi_2\overline{\psi}_2),\ee
and implies (\ref{b12}). It may be worth noting that (\ref{b13}) also implies
\[J=-3\psi_2\overline{\psi}_2(\psi_2+\overline{\psi_2}),\]
which implies the reality of $J$. 
We claim that the condition (\ref{b13}) (and therefore (\ref{b12})) is not new, but follows from the reality of $(I, J)$. To see it
consider a combination $2J+3I\psi_2$, where $(I, J)$ are given by
(\ref{i1}) with $\psi_0=\psi_4=0$. This gives  
\be
\label{cubicap}
6(\psi_2)^3-3I\psi_2-2J=0.
\ee
Taking the imaginary part of (\ref{cubicap}), and using reality
of $(I, J)$ gives
\[
2[(\psi_2)^3-(\bar{\psi}_2)^3]=I(\psi_2-\bar{\psi}_2).
\]
In Case (b) we have $\psi_2$ not real. We can therefore divide both sides
of this equation by $\psi_2-\bar{\psi}_2$ which gives (\ref{b13}).
\medskip
We still have to impose the real part of (\ref{b3}). Using (\ref{b13}) this (eventually) is
\[(\psi_2-\overline{\psi_2})^4=16(\psi_1\psi_3+\overline{\psi}_1\overline{\psi}_3)^2,\]
which is already known from (\ref{b11a}). The solution for $\sigma_{ABA'B'}$ is essentially unique, up to choices of $\epsilon$ and $r$.
\subsection*{Summary}
Assuming the reality of $I$ and $J$ we were able to show that real solutions
to (\ref{em4}) always exists in algebraic types N, III, D and II. Type 
I was more complicated. We can always make a 
choice $\psi_0=\psi_4=0$, which does not alter the reality of $(I, J)$. The 
analysis then branches: if $\psi_2$ is real, then reality of $I$
forces $\psi_1\psi_3$ to be real, and solutions to (\ref{em4}) exist.
If $\psi_2$ is not real, then reality of $(I, J)$ imply
a condition (\ref{b13}) which is sufficient for solutions to
(\ref{em4}) to exist.
\koniec
\section{The GHP formalism}
In this appendix we shall summarise the weighted form of the Newman-Penrose formalism developed by
Geroch,  Held, and Penrose  \cite{ghp}.
One begins with a choice of normalised spinor dyad:
\[(o^A,\iota^A)\mbox{  with  }o_A\iota^A=1.\]
In the case of the Schwarzschild metric these can conveniently be taken to be the Principal Null Directions of the curvature, and in any spherically symmetric metric they can be taken to be associated with the radially outgoing and 
radially ingoing null directions. There is the freedom to change the dyad according to
\[(o^A,\iota^A)\rightarrow(\tilde{o}^A,\tilde{\iota}^A)=(\lambda o^A,\lambda^{-1}\iota^A),\]
and one associates with this freedom the notion of \emph{GHP weight}: a space-time scalar $\eta$ has GHP weight $(p,q)$ if it transforms as
\[\eta\rightarrow\tilde{\eta}=\lambda^p\overline{\lambda}^q\eta\]
under this rescaling. GHP weights are related to the earlier notions (see \cite{ghp} or \cite{pr}) of \emph{spin weight} and \emph{boost weight}, respectively $s$ and $w$, according to
\[s=\frac12(p-q),\;\;\;w=\frac12(p+q).\]

\medskip

As in the standard Newman-Penrose formalism, one introduces a null tetrad according to
\[\ell^a=o^A\overline{o}^{A'},\;\;n^a=\iota^A\overline{\iota}^{A'},\;\;m^a=o^A\overline{\iota}^{A'},\;\;\overline{m}^a=\iota^A\overline{o}^{A'},\]
and these vectors have GHP weights $(1,1),(-1,-1),(1,-1)$ and $(-1,1)$ respectively. The formalism admits a symmetry conveniently called `priming' according to which
\[(o^A,\iota^A)\rightarrow(i\iota^A,io^A)\]
and then for example
\[(\ell^a)'=n^a,\;\;(n^a)'=\ell^a,\;\;(m^a)'=\overline{m}^a,\;\;(\overline{m}^a)'=m^a.\]
It can be checked that the prime $\eta'$ of a quantity $\eta$ with GHP weight $(p,q)$ has weight $(-p,-q)$ and 
\[(\eta')'=(-1)^{p+q}\eta,\]
so that prime is nearly an involution.

One labels the tetrad components of the gradient as in the NP formalism:
\[D=\ell^a\nabla_a,\;\;\Delta=n^a\nabla_a,\;\;\delta=m^a\nabla_a,\;\;\overline{\delta}=\overline{m}^a\nabla_a,\]
but these operators don't have good GHP weight. We shall modify them shortly.

The NP spin coefficients are the components of the spin connection in the chosen dyad according to the scheme
\[Do^A=\epsilon o^A-\kappa\iota^A,\;\;\;\;D\iota^A=\pi o^A-\epsilon \iota^A,\]
\[\Delta o^A=\gamma o^A-\tau\iota^A,\;\;\;\;\Delta\iota^A=\nu o^A-\gamma\iota^A,\]
\[\delta o^A=\beta o^A-\sigma\iota^A,\;\;\;\;\delta\iota^A=\mu o^A-\beta\iota^A,\]
\[\overline{\delta}o^A=\alpha o^A-\rho\iota^A,\;\;\;\;\overline{\delta}\iota^A=\lambda o^A-\alpha\iota^A.\]

Eight of the spin coefficients have good GHP weight and they are related in pairs by prime so that it becomes convenient to eliminate four of them as primes of four others. These are
\[\mu=-\rho',\;\;\;\;\lambda=-\sigma',\;\;\;\;\pi=-\tau',\;\;\;\;\nu=-\kappa'.\]
The other four spin coefficients are conveniently incorporated into weighted operators, thorn and edth, according to 
\be\label{g1}\th=D-p\epsilon-q\overline{\epsilon},\;\;\;\;\et=\delta-p\beta-q\overline{\alpha},\ee
when acting on weight $(p,q)$ quantities, together with their primes
\be\label{g2}\th'=\Delta-p\gamma-q\overline{\gamma},\;\;\;\;\et'=\overline{\delta}-p\alpha-q\overline{\beta}.\ee

\medskip

With the metric (\ref{sc1}) we choose the NP tetrad to be
\[D=\frac{1}{F}\partial_u,\;\;\Delta=\frac{1}{F}\partial_v,\;\;\delta=\frac{P}{r\sqrt{2}}\partial_\zeta,\]
and there are simplifications: the only nonzero spin-coefficients are 
\[\rho=-\frac{r_u}{Fr},\;\;\rho'=-\frac{r_v}{Fr},\;\alpha=-\overline{\beta}=\overline{\alpha}'=\frac{\zeta}{2\sqrt{2}r},\;\;\epsilon=\frac{F_u}{2F^2},\;\;\gamma=-\epsilon'=-\frac{F_v}{2F^2},\]
where one may substitute for $r_u,r_v,F_u,F_v$ if desired. The Ricci tensor is zero and the only nonzero component of the Weyl spinor is $\psi_2=-m/r^3$. The weighted derivatives of the dyad simplify to
\[\th o^A=\th' o^A=\th \iota^A=\th'\iota^A=0=\et o^A=\et'\iota^A\]
and
\[\et'o^A=-\rho\iota^A,\;\;\;\;\et\iota^A=-\rho' o^A.\]
The commutators of the GHP operators are given in \cite{ghp} in the general case but here they simplify to
\be\label{comm1}[\th,\et]=\rho\et,\;\;\;[\th',\et]=\rho'\et,\ee
with their primes, and 
\be\label{comm2}[\et,\et']=(p-q)(\rho\rho'+\psi_2),\;\;[\th,\th']=-(p+q)\psi_2,\ee
where one calculates
\[\rho\rho'+\psi_2= -\frac{1}{2r^2} \]
which is minus half the Gauss curvature of the constant $r$ sphere.

Spherical symmetry of the metric and tetrad implies that $\rho,\rho'$ and $\psi_2$ are spherically symmetric in that 
\[\et\rho=\et\rho'=\et\psi_2=0=\et'\rho=\et'\rho'=\et'\psi_2,\]
which we use in the proof of Theorem \ref{theop}.
\section{Infinite boost limit of the Kasner embedding}
\setcounter{equation}{0}
\def\theequation{\thesection\arabic{equation}}
In \cite{japanese} the authors have systematically analysed isometric embeddings of the metric
(\ref{four_m}) with $V=1-2m/r$ in $\R^{5,1}$ and $\R^{4,2}$ with a metric\footnote{In this appendix we shall use the $(- + + +)$ signature, 
so that our results agree with \cite{japanese}.}
\be
\label{ds22}
ds^2=-d{Z_0}^2+\epsilon d{Z_1}^2+d{Z_2}^2+ d{Z_3}^2+d{Z_4}^2+d{Z_5}^2
\ee
such that 
\[
Z_2= r\sin{\theta}\sin{\phi}, \quad Z_3= r\sin{\theta}\cos{\phi}, \quad Z_4=r\cos{\theta}.
\]
The problem then reduces to finding all isometric embeddings of a metric
$\gamma=-Vdt^2+(V^{-1}+1)dr^2$ into $\R^{2,  1}$ or $\R^{1, 2}$, and the general solution
was obtained under an additional assumption that the second fundamental form of the embedding is
diagonal. There are three embedding types which can be classified according to 
whether
trajectories of $\p/\p t$ lift to circles (elliptic, Kasner-type embedding)
with $\epsilon=-1$, hyperbolae
(hyperbolic, Fronsdal type embedding with $\epsilon=1$) or parabolas with $\epsilon=-1$. 

We shall demonstrate that the parabolic embedding can be obtained by taking a  
Lorentz boost
of the elliptic isometric embedding, relating the boost parameter $c$ to 
the mass in the Schwarzschild solution by $c=1/4m$ (or more generally to the surface gravity
in case of the general $V(r)$), and taking a limit $c\rightarrow 0$. We shall just
give the embedding forms $Z_0(t, r), Z_1(t, r), Z_5(t, r)$ which have been obtained for any $V(r)$
by trial and error.
\begin{eqnarray*}
Z_5&=&\frac{(1+c^2)(c^2 L-\sqrt{V})+(1-c^2)\sqrt{V} \cos{(ct)}}{2c^2}\\
Z_0&=&\frac{(1-c^2)(c^2 L-\sqrt{V})+(1+c^2)\sqrt{V} \cos{(ct)}}{2c^2}\\
Z_1&=& \frac{\sqrt{V} \sin{(ct)}}{c}, \quad\mbox{where}\quad V=V(r), L=L(r).
\end{eqnarray*}
Then $ds^2$ given by (\ref{ds22}) with $\epsilon=-1$ pulls back to 
\[
g=-V dt^2+ \frac{L'(m^2VL'-V'\sqrt{V})+V}{V}dr^2 +r^2(d\theta^2+\sin{\theta}^2d\phi^2),
\]
and to recover the metric (\ref{four_m}) the function $L(r)$ must satisfy a 1st order ODE
\[
(m^2VL'-V'\sqrt{V})L'+V-1=0.
\]
In the limit $c\rightarrow 0$ we get  a parabolic embedding
\[
Z_5=-\sqrt{V}+\frac{1}{2}L-\frac{t^2}{4} \sqrt{V}, \quad
Z_0=\sqrt{V}+\frac{1}{2}L-\frac{t^2}{4} \sqrt{V}, \quad Z_1=t\sqrt{V}
\]
and
\[
g=-Vdt^2 +\frac{V-L'V'\sqrt{V}}{V}dr^2+r^2(d\theta^2+\sin{\theta}^2d\phi^2),
\]
where 
\[
L(r)=\int^r \frac{V(\rho)-1}{V'(\rho)\sqrt{V(\rho)}}d\rho.
\]
In the case of the Schwarzschild metric this reduces to the quadratic embedding of \cite{japanese} (see also \cite{pestun}).
\section{The Fronsdal embedding from the conformal embedding}
\setcounter{equation}{0}
\def\theequation{\thesection\arabic{equation}}
\label{sec_frons}
We shall now show how the the Fronsdal \cite{fronsdal} isometric
embedding $\phi:M\rightarrow \R^{5, 1}$ with
\[
g=\phi^*(d{Z_0}^2-d{Z_1}^2-\dots-d{Z_5}^2) 
\]
is related to our embedding (\ref{cembedding}). 
Set
\[
Z_\mu=\Omega^{-1} X_\mu  \quad\mbox{for} \quad\mu=0, \dots, 4, \quad \mbox{and}\quad 
Z_5=Z(r)
\]
and compute
\begin{eqnarray*}
g&=&\eta^{{\mu}{\nu}} dZ_\mu dZ_\nu-dZ^2\\
&=&\Omega^{-2} (\eta^{{\mu}{\nu}} dX_\mu dX_\nu+K(\Omega^{-1}d\Omega)^2-\Omega^{-1}d\Omega dK) -dZ^2
\end{eqnarray*}
where
\[
K\equiv\eta^{{\mu}{\nu}} X_\mu X_\nu=\Omega^2\frac{32m^3 -16m^2r-r^3}{r}.
\]
Therefore $Z(r)$ is determined by
\be
\label{Zprime}
\Omega^2(Z')^2=K[(\ln{\Omega})']^2-K'(\ln{\Omega})'.
\ee
Using $\Omega=h(r)/r$ and (\ref{eq7}) we find
\[
(\ln{\Omega})'=\frac{p}{q}-\frac{1}{r},  \quad K=\Omega^2\frac{q}{r^2},
\]
where $(p, q)$ are given by (\ref{pq}). Substituting this into (\ref{Zprime})
gives
\[
Z_5=\int\sqrt{\frac{2m}{r}+ \Big(\frac{2m}{r}\Big)^2+
\Big(\frac{2m}{r}\Big)^3}dr
\]
in agreement with \cite{fronsdal}.
The diagrams below show the radial embeddings with $\theta=\phi=\mbox{const}$
projected to a three-dimensional space with coordinates $Z_0=\Omega^{-1} T, Z_1=\Omega^{-1} X,
Z_5$.
 \begin{center}
  \label{fronsdal_graphs}
  \includegraphics[width=5cm,height=5cm,angle=0]{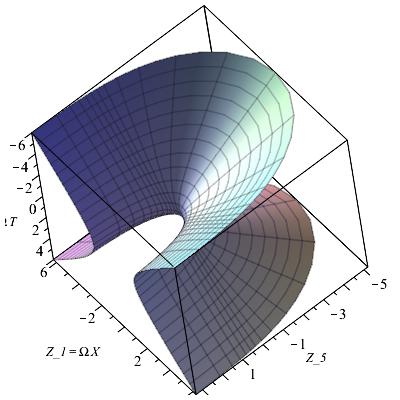}, \;\;
  \includegraphics[width=5cm,height=5cm,angle=0]{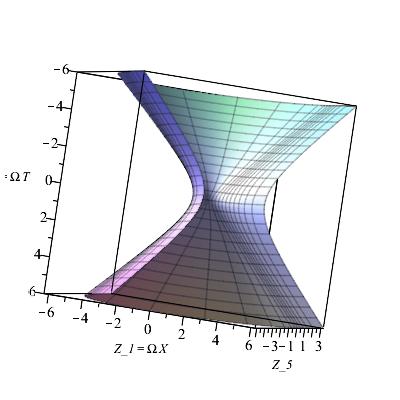}.
\begin{center}
{{\bf Figure 4.} The radial Fronsdal embedding.}
\end{center}
\end{center}
\end{appendix}

\end{document}